\documentclass[12pt,a4paper]{article}

\usepackage[font=small]{caption}
\usepackage[text={470pt,650pt},headheight={14.5pt},centering]{geometry}
\usepackage{amssymb,amsmath,color,graphics,amscd,epsf,indentfirst,amsfonts}
\usepackage{mathtools}
\usepackage{enumerate}
\usepackage{epsfig}
\usepackage[colorlinks=true,citecolor=blue]{hyperref}
\usepackage{cite}
\usepackage{nicefrac}
\usepackage{scalefnt}
\usepackage[titles]{tocloft}
\usepackage{sectsty}
\allsectionsfont{\bf \scalefont{.7} \selectfont}
\subsectionfont{\bf \scalefont{.85} \it \selectfont}
\subsubsectionfont{\bf \scalefont{1} \it \selectfont}
\usepackage[T1]{fontenc}
\usepackage{lmodern}
\usepackage{bibspacing}

\newcommand{\otau}{\overline{\tau}}

\newcommand{\hhbar}{\relax}

\makeatletter
\renewcommand{\@dotsep}{4.5}
\makeatother

\def\be{\begin{equation}}
\def\ee{\end{equation}}

\makeatletter
\def\@seccntformat#1{\csname the#1\endcsname.\quad}
\makeatother

\def\clock{{\count0=\time
           \divide\count0 60
           \ifnum\count0<10 0\fi\the\count0
           \multiply\count0 -60 \advance\count0 \time
           :\ifnum\count0<10 0\fi \the\count0
         }}
\newcommand{\timestamp}{{\small\vbox{\hbox{\tt\jobname.tex}
\hbox{\the\day/\the\month/\the\year, \clock}}}}

\def\AA{{\cal A}}
\def\BB{{\cal B}}
\def\CC{{\cal C}}
\def\DD{{\cal D}}

\def\HH{{\cal H}}

\def\JJ{{\cal J}}

\def\LL{{\cal L}}
\def\MM{{\cal M}}
\def\NN{{\cal N}}
\def\OO{{\cal O}}

\def\QQ{{\cal Q}}
\def\RR{{\cal R}}
\def\SS{{\cal S}}
\def\TT{{\cal T}}

\def\IR{{\mathbb R}}
\def\IT{{\mathbb T}}
\def\IZ{{\mathbb Z}}

\def\tts{{$tt^*$ }}

\def\Tr{{\rm {Tr}}}

\def\d{{\partial}}
\def\p{{\partial}}
\def\time{{\tau}}

\def\beq{\begin{equation}}
\def\eeq{\end{equation}}
\newcommand{\bea}{\begin{eqnarray}}
\newcommand{\eea}{\end{eqnarray}}
\def\bal{\begin{align}}
\def\eal{\end{align}}

\def\bF{{\bf F}}
\def\bA{{\bf A}}
\def\g{e}
\def\au{\Theta}

\numberwithin{equation}{section}

\begin{document}
\begin{titlepage} 
\begin{flushright}
CERN-TH-2016-241\\
DCPT-16/55
\vskip -1cm
\end{flushright}
\vskip 1.9cm
\begin{center}
\font\titlerm=cmr10 scaled\magstep4
    \font\titlei=cmmi10 scaled\magstep4
    \font\titleis=cmmi7 scaled\magstep4
    \centerline{\LARGE \titlerm 
    Aspects of Berry phase in QFT}
\vskip 1.5cm
{Marco Baggio,$^a$ Vasilis Niarchos,$^b$ Kyriakos Papadodimas$^{c,d}$} \\
\vskip 0.5cm
{\it $^a$Institute for Theoretical Physics, KU Leuven}\\
{\it Celestijnenlaan 200D, B-3001 Leuven, Belgium}\\
\medskip
{\it $^b$Department of Mathematical Sciences and Center for Particle Theory}\\
{\it Durham University, Durham, DH1 3LE, UK}\\
\medskip
{\it $^{c}$Theory Group, Physics Department, CERN, CH-1211 Geneva 23, Switzerland}\\
{\it $^{d}$Van Swinderen Institute for Particle Physics and Gravity, University of Groningen, Nijenborgh 4,
9747 AG Groningen, The Netherlands}\\
\medskip
{$^a$marco.baggio@kuleuven.be, $^b$vasileios.niarchos@durham.ac.uk}\\
{$^{c,d}$kyriakos.papadodimas@cern.ch}

\end{center}
\vskip .2cm
\centerline{\bf Abstract}

\vskip .2cm 
\noindent
When continuous parameters in a QFT are varied adiabatically, quantum states typically 
undergo mixing---a phenomenon characterized by the Berry phase.
We initiate a systematic analysis of the Berry phase in QFT using standard quantum mechanics methods.
We show that a non-trivial Berry phase appears in many familiar QFTs. We study a variety of examples including free electromagnetism with a theta angle, and certain supersymmetric QFTs in two and four spacetime dimensions. We also argue that a large class of QFTs with rich Berry properties is provided by CFTs with non-trivial conformal manifolds. Using the operator-state correspondence we demonstrate in this case that the Berry connection is equivalent to the connection on the conformal manifold derived previously in conformal perturbation theory. In the special case of chiral primary states in 2d ${\cal N}=(2,2)$ and 4d ${\cal N}=2$ SCFTs the Berry phase is governed by the $tt^*$ equations. We present a technically useful rederivation of these equations using quantum mechanics methods.

\vfill
\noindent

\end{titlepage}\vfill\eject

\hypersetup{pageanchor=true}

\setcounter{equation}{0}

\pagestyle{empty}
\small
\vspace*{-0.7cm}
{
\hypersetup{linkcolor=black}
\tableofcontents
}
\normalsize

\pagestyle{plain}
\setcounter{page}{1}
 
\newpage
\section{Introduction} 
\label{setup}

When the Hamiltonian of a quantum system depends on continuous parameters, a natural connection can be defined on the parameter space. This is called  Pancharatnam-Berry connection \cite{pancha, Berry:1984jv} and it encodes the geometric phase that quantum states pick up under adiabatic variations of the parameters. It has played an important role in several physical systems in condensed matter and atomic physics,  see \cite{wilczek1989geometric,chruscinski2004geometric} for a review.

In this paper we present new results about the Berry phase in quantum field theory (QFT). As in any other quantum system, when we adiabatically vary the parameters of the Lagrangian, quantum states in the Hilbert space will generically pick up a non-trivial Berry phase.
We show that a non-trivial Berry phase can be encountered even in simple weakly coupled quantum field theories. At strong coupling it is typically very hard to compute the Berry phase analytically. However, as we demonstrate, in supersymmetric QFTs in various dimensions exact results about the Berry phase of special (BPS) states can be derived, which hold for all values of the coupling.

In QFT the validity of the adiabatic theorem and the precise computation of the Berry phase may be subtle in situations with continuous spectra. We will deal with such issues by placing the theory on a spatial compact manifold, e.g.\ the torus, the sphere etc. Certain manifolds may be more convenient than others for the study of the Berry phase of specific states.
In many of the examples that we study in this paper this approach allows us to draw conclusions about the QFT in flat space. In some cases we can obtain a sensible decompactification limit, where the results are insensitive to the specific details of the compact manifold. In the case of conformal field theories, the relation to the theory in flat space is achieved by a conformal transformation and the operator-state correspondence.

A simple example of a weakly coupled quantum field theory that exhibits a non-trivial Berry phase is four-dimensional pure electromagnetism in the presence of a theta angle. The Hilbert space of this theory is a freely generated Fock space of photons. Even though the theory is free and the eigenvalues of the Hamiltonian do not depend on the coupling constant $\g$ and theta angle $\theta$, the actual eigenstates do depend on these parameters. As a result, we show that the states of the theory exhibit non-trivial Berry phase under adiabatic variations of $\g$ and $\theta$. In terms of the complexified gauge coupling $\tau = \frac{\theta}{2\pi} + i \frac{4\pi}{\g^2}$ we find that a state with ${\mathfrak n}_+$ (${\mathfrak n}_- $) photons of positive (negative) helicity has an associated Berry curvature given by
\beq
\label{photonadintro}
\bF^{({\mathfrak n}_+,{\mathfrak n}_-)}_{\tau \overline{\tau}} = \frac{1}{8} ({\mathfrak n}_+ - {\mathfrak n}_-) \frac{1}{({\rm Im}\tau)^2}
~.
\eeq
In particular, this formula implies that if a photon propagates in a medium with a slow variation of the effective $\g,\theta$ couplings, its polarization vector will undergo a spatial rotation. Such effects might be visible in suitable setups of topological insulators.

This result generalizes very naturally to the low energy $U(1)^r$ theory that characterizes the Coulomb branch  of four-dimensional ${\cal N}=2$ theories.
In that case we argue that the low energy photons have a non-abelian $U(r)$ Berry curvature that is proportional to the curvature of the Seiberg-Witten metric on the Coulomb branch \cite{Seiberg:1994rs}.

In sections \ref{massive}, \ref{2d}, \ref{4d2} we present several new computations of the Berry connection for certain states in supersymmetric quantum field theories. In continuous families of supersymmetric theories, the deformations of the Hamiltonian are often $Q$-exact and related to $F$-terms. This leads to drastic simplifications in the computation of the Berry curvature of supersymmetric ground states and, more generally, of BPS or ``chiral'' states.
We work out the case of massive ${\cal N}=1$ theories in four dimensions and show that the Berry curvature of supersymmetric ground states takes a particularly simple form. In our second example we consider the case of chiral primary states in the NS sector of 2d ${\cal N}=(2,2)$ SCFTs on $\IR \times S^1$.
We show that the Berry phase of these states is governed by the $tt^*$ equations, reproducing the classic results of \cite{Cecotti:1991me} and \cite{deBoer:2008ss}. The new derivation demonstrates how the $tt^*$ equations arise from a straightforward manipulation of the Berry curvature formula bypassing the use of the superconformal Ward identities needed in conformal perturbation theory \cite{deBoer:2008ss}. Similarly, by analyzing 4d ${\cal N}=2$ SCFTs on $\IR \times S^3$ we prove that the Berry curvature of $\NN=2$ chiral primary states is governed by the 4d analogue of the $tt^*$ equations derived in \cite{Papadodimas:2009eu}.

An important consequence of our analysis is that the Berry curvature can be computed \emph{exactly} in various 4d ${\cal N}=2$ SCFTs. For example, the results of \cite{Baggio:2014sna,Baggio:2014ioa}, where the exact three-point functions of chiral primary operators were determined for $SU(2)$ $\mathcal{N}=2$ superconformal QCD, can now be interpreted as providing the exact Berry curvature of the chiral primary states of the theory.

The above results in 2d $\NN=(2,2)$ and 4d $\NN=2$ SCFTs are examples of a more general relation between the Berry connection for states of a CFT on $\IR \times S^{d-1}$ and the connection on the space of operators that is naturally defined in conformal perturbation theory \cite{Kutasov:1988xb, Ranganathan:1993vj}. In section \ref{specVScft} we present a general formal argument based on the operator-state correspondence that exhibits the equivalence of the two connections in any CFT with a non-trivial conformal manifold and for any set of states/operators. In particular, we arrive at a very natural physical interpretation for the curvature of the Zamolodchikov metric: it characterizes the Berry phase of the marginal operators as the marginal parameters undergo an adiabatic cyclic variation.

Previous works which considered the Berry phase in systems with supersymmetry include \cite{Sonner:2008fi,Laia:2010dv} and references therein.

\section{Review of Berry phase}
\label{BerryReview}

In this section we provide a lightning review of basic properties of the Berry phase in quantum mechanics that will be
useful in this paper. Extended reviews of the subject can be found, for example, in \cite{wilczek1989geometric,chruscinski2004geometric}.

Consider a Hamiltonian that depends on a set of continuous parameters $\lambda^i$, where $i=1,..,k$. We think of the parameter space as a $k$-dimensional manifold ${\cal M}$.
For now we assume that these parameters are real numbers. In supersymmetric theories it is more natural to combine them into complex combinations and the parameter space may be a complex manifold.

Let us further assume that there is a fixed Hilbert space $\HH$ where the Hamiltonian $H(\lambda^i)$ acts in a prescribed $\lambda$-dependent manner. 
We assume that this Hilbert space exists at least for local patches on the manifold $\MM$.
For every choice of the parameters $\lambda^i$, there is a basis of eigenvectors of the Hamiltonian, which will be denoted as $|n(\lambda)\rangle$. By definition,
\beq
\label{revaa}
H(\lambda) |n(\lambda)\rangle = E_n(\lambda) |n(\lambda)\rangle
~.
\eeq
For starters, let us consider the case of a non-degenerate spectrum over a region of the parameter space, thus excluding the possibility of level-crossing. The case of degenerate spectra will be discussed in the next subsection.
In the absence of degenerate spectra the eigenvectors of the Hamiltonian are uniquely fixed up to a phase at any given value of the parameters. The physics of this phase can exhibit interesting effects.

\subsection{Abelian Berry connection}

\noindent {\bf Definition.} 
We select the eigenvectors $|n(\lambda)\rangle$ over a region of the parameter space with an {\it arbitrary} 
$\lambda$-dependent choice of phase.
Following Berry \cite{Berry:1984jv} we define the object\footnote{Typically the Berry connection is defined as a real object with an overall $i$ factor in \eqref{berrycon}. To conform with standard definitions of the connection in conformal perturbation theory later on, we will not include the factor of $i$ in the present definitions.}
\be
\label{berrycon}
\bA^{(n)}_i = \langle n(\lambda)|  \frac{\partial}{\partial \lambda_i} |n(\lambda)\rangle~.
\ee
This should be understood in geometric terms as a {\it connection} encoding how to compare the phase of the state $|n\rangle$ at nearby points of the parameter space $\MM$. Notice that, since we assumed no level crossings, we can unambiguously keep track of the state $|n\rangle$ as we move on the parameter space. Moreover, under a change of the choice of phase of the eigenvectors, $|n'(\lambda)\rangle = e^{i \phi^{(n)}(\lambda)} |n(\lambda)\rangle$, the object \eqref{berrycon} transforms as a $U(1)$ gauge field
\beq
\label{revab}
\bA^{{(n)}'}_i =  \langle n(\lambda)|  e^{-i\phi^{(n)}(\lambda)}\frac{\partial}{\partial \lambda_i}  \left(e^{i\phi^{(n)}}|n(\lambda)\rangle\right) = \bA^{(n)}_i + i \partial_i \phi^{(n)}
~.
\eeq
We have a $U(1)$ gauge field for each state $|n\rangle$.

\vskip10pt
\noindent {\bf Berry curvature.}
The field strength of this gauge field (equivalently, the curvature of the above connection) has components
\beq
\label{revac}
\bF^{(n)}_{ij} \equiv \partial_i \bA^{(n)}_j - \partial_j \bA^{(n)}_i
~.
\eeq
After a few standard manipulations (see appendix \ref{spectralBerry}) we find that the curvature can be expressed as a spectral sum
\be
\label{berrycurv}
\bF^{(n)}_{ij}  = \sum_{m \neq n} \frac{\langle n| \partial_i H |m\rangle \langle m| \partial_j H |n\rangle - (i \leftrightarrow j)}{(E_m-E_n)^2}~.
\ee
The intermediate states $|m\rangle$ are assumed orthonormal.
As expected, the formula for the curvature is invariant under a change in the choice of phases of the wavefunction.

\vskip10pt
\noindent {\bf Mathematical structure.} The parameter space ${\cal M}$ and the complex line $\HH_n$ (representing the Hilbert space at energy $E_n$) over each point of $\MM$ defines a vector bundle over the parameter space. A further restriction on normalized states in $\HH_n$ defines a principal $U(1)$ fiber bundle with $\MM$ as the base space.
These bundles are equipped with a natural connection \cite{Simon:1983mh}, which coincides with the connection computed by \eqref{berrycon}. A choice of eigenvectors $|n(\lambda)\rangle$ corresponds to a family of sections in this bundle. 

\vskip10pt

\noindent {\bf Physical origin of the Berry connection.}
The Berry connection is of course related to the adiabatic theorem of quantum mechanics. If we start with the system in one of the energy eigenstates $|n(\lambda)\rangle$ and change the parameters of the Hamiltonian very slowly, the system evolves by remaining in the instantaneous eigenstate $|n(\lambda(t))\rangle$. Apart from the trivial ``dynamical phase'', which is $e^{-i E_n t}$, the Schroedinger evolution of the state also picks up an additional ``geometric phase''. Berry discovered that in cyclic variations this phase is a characteristic quantity of the system that depends only on the path taken on the parameter space. It is given by integrating the connection \eqref{berrycon} along the path. In this sense, the dynamics of quantum mechanics selects a particular connection.

\subsection{Non-abelian generalization}

If there is a subspace of degenerate states (and the degeneracies are not accidental, but rather persist over an open set on the parameter space), then the Berry connection may be non-abelian \cite{Wilczek:1984dh}.
If $N$ is the degeneracy of an energy subspace $E_n$, then the connection is generally in the adjoint representation of $U(N)$. Let us select an arbitrary basis for the degenerate states of energy $E_n$ as $|n,a (\lambda)\rangle$, where $a=1,...,N$. Then, the formulae for the connection and curvature become
\beq
\label{revba}
(\bA^{(n)}_{i})_{ab} = \langle n,b| \partial_i |n,a \rangle 
~,
\eeq
with curvature
\beq
(\bF^{(n)}_{ij})_a^{~b} = \partial_i (\bA^{(n)}_{j})_{a}^{~b} - \partial_j (\bA^{(n)}_{i})_{a}^{~b} - [\bA^{(n)}_i,\bA^{(n)}_j]_{a}^{~b}
~.
\eeq
The indices $a,b$ are raised and lowered with the 2-point function 
\beq
\label{revbc}
g_{(n)ab} = \langle n, a | n, b \rangle
~.
\eeq
The non-abelian generalization of eq.\ \eqref{berrycurv} reads
\beq
\label{revbb}
\left( \bF^{(n)}_{\mu\nu} \right)_{ab}
=  \sum_{m \neq n} \sum_{c,d} 
\frac{1}{(E_n-E_m)^2} \langle n,b | \p_\mu H | m,c \rangle g_{(n)}^{cd} \langle m,d | \p_\nu H | n,a \rangle 
- (\mu \leftrightarrow \nu)
~.
\eeq
In this formula, whose derivation is summarized in appendix \ref{spectralBerry} for the benefit of the reader, the spectral sum is performed over a general intermediate complete basis of states (not necessarily orthonormal) with overlap $g_{(n)ab}$.

\subsection{Systems with time-reversal or CPT invariance} 

If a quantum system is invariant under an anti-unitary symmetry $\au$, for example time-reversal or CPT, then the Berry phase is constrained. The anti-linear operator $\au$ obeys $\au^\dagger \au=1$ and
\beq
\label{symmetry}
[H(\lambda),\au] = 0\qquad \forall \lambda
~.
\eeq
For a non-degenerate energy eigenstate it is easy to prove that the anti-unitary symmetry implies a vanishing Berry phase.
One general implication of this result is the following. Relativistic QFTs are invariant under CPT. If in addition the ground state is unique, then the Berry phase associated to it should be zero. Notice that this result holds even for a QFT defined on a manifold of the form $\IR \times {\cal T}$, provided that CPT remains true and that the ground state is unique. 

In the non-abelian case, the anti-unitary symmetry implies that the Berry connection reduces from $U(N)$ down to $SO(N)$, in the case that $\au^2=1$, and down to $Sp(N)$ if $\au^2=-1$. We present a proof of these statements in appendix \ref{antiproof}.

\subsection{Berry phase in QFT}

Quantum field theories are typically quantum systems that depend on a number of continuous parameters, such as masses and other couplings. We want to understand how to compute the Berry phase of various states in QFT associated with the adiabatic change of such parameters.
Since a QFT is a quantum system with an infinite number of degrees of freedom, one may face infrared (IR) and/or  ultraviolet (UV) problems in applying the previous formulae. One of the possible IR subtleties that can arise if a QFT is defined in infinite flat space is the appearance of a continuous spectrum. In that case extra care needs to be taken with the normalization of the states as well as the applicability of the adiabatic theorem. 
Typically, these subtleties can be avoided by formulating the QFT on a spatial compact manifold $\TT$. The QFT on the hypercylinder $\IR \times \TT$ is essentially quantum mechanics with a complicated infinite tower of states. Different compact manifolds $\TT$ can lead to different Hamiltonians, and some choices of $\TT$ may be more preferable for a specific set of questions compared to others.
In the following sections we will see examples where it is more convenient to compactify a QFT on a spatial torus. In other situations (most notably CFTs) the natural IR regularization occurs on a round sphere. On the other hand, issues related to UV divergences can be dealt with using standard methods of renormalization.

Modulo the above potential subtleties, it is in principle conceptually straightforward to compute the Berry phase of any state in QFT. The Hamiltonian, as a function of external couplings, is derived in the canonical formalism and the Berry connection/curvature is evaluated using the formulae \eqref{revba}, \eqref{revbb}. Typically, this will lead to a rather involved computation where analytic, closed form results will be out of reach.
In weakly coupled theories one could proceed with perturbative/path integral methods. There can be, however, special situations where the symmetries of the underlying theory allow more powerful, even non-perturbative, results. Supersymmetric QFTs provide such examples.

\section{A warmup example: free electromagnetism}

Berry phases can appear even in basic QFTs. To illustrate this point, in this section we consider a $U(1)$ gauge theory with coupling constant $\g$.
We also introduce the $\theta$-angle and consider the Lagrangian
\beq
\label{lagqed}
\LL = -\frac{1}{4 \g^2} F_{\mu\nu} F^{\mu\nu} + \frac{\theta}{32\pi^2} F_{\mu\nu} \widetilde{F}^{\mu\nu}~,
\eeq
where
$\widetilde{F}_{\mu\nu} = \frac{1}{2} \epsilon_{\mu\nu\rho\sigma}F^{\rho\sigma}$. When $\theta$ is constant the $\theta$-interaction is a total derivative that does not affect the classical dynamics of the theory. However, $\theta$ can have important physical implications in the presence of magnetic monopoles, nontrivial cycles in the geometry, boundaries or interfaces where gradients of $\theta(x)$ appear (see for example \cite{Wilczek:1987mv}).
As we shall see, it also leads to a nontrivial Berry phase.

In this section we are interested in the properties of photon states as we vary adiabatically both couplings $\g$ and $\theta$. Hence, the parameter space ${\cal M}$ in this context can be thought of as the upper half plane parametrized by $\tau = \frac{\theta}{2\pi} + i \frac{4\pi}{\g^2}$, modulo global identifications.
Defining
\beq
\label{lagqedc}
F_{\pm, \mu\nu}  = F_{\mu\nu} \pm \frac{1}{2}i \epsilon_{\mu\nu\rho \sigma} F^{\rho \sigma}
~,
\eeq
the Lagrangian \eqref{lagqed} can be written more explicitly in terms of $\tau$ (and its complex conjugate) as
\beq
\label{lagqedd}
\LL = \frac{i}{64 \pi} \tau F_+^2  - \frac{i}{64 \pi } \overline{\tau} F_-^2
~.
\eeq
We remind the reader that there is a natural metric on the parameter space $\MM$ of the form
\beq
\label{hypmet}
ds^2 = \frac{d\tau d\overline{\tau}}{({\rm Im} \tau)^2}~.
\eeq

The Hilbert space of electromagnetism consists of the vacuum as well as states with an arbitrary number of free photons. In this case the spectrum of energy eigenstates does not depend on the couplings $\g,\theta$. 
This does not mean, however, that the eigenvalue problem is independent of $\tau$. While the eigenvalues of the Hamiltonian do not depend on $\tau$, the {\it eigenvectors} do rotate inside the Hilbert space when we vary the couplings $\g,\theta$. For that reason we get a nonzero Berry phase even for a free $U(1)$ theory.

As we mentioned already in the previous section, if we define the theory on ${\mathbb R}^{1,3}$, the notion of individual photons becomes somewhat subtle because of the infinite volume. In particular, the energy eigenstates correspond to momentum eigenstates that are only $\delta$-function normalizable. Computing the Berry phase for such states may require extra care. Moreover, in the expression \eqref{berrycurv}, we should exercise appropriate caution when dealing with a continuous spectrum of intermediate states $|m\rangle$ and the exclusion rule $E_m\neq E_n$. These are obviously IR issues that can be dealt with if we define the theory on a compact manifold.
For example, we can define the theory on a spatial torus or a sphere. In this section we choose the torus.  We will find that the Berry phase picked up by quantum states is independent of the volume of the torus. In addition, one can check that under a general (not necessarily adiabatic) time-dependent variation of the couplings the total time derivative of the Hamiltonian does not induce mixing between photon states of different frequencies.
These observations imply that the computed phase will persist in the infinite volume limit, and that there are no subtleties in the adiabatic limit as the volume of the torus becomes larger and larger.

\subsection{Berry phase of photon states}

We can compute the Berry phase by a straightforward application of the formula \eqref{berrycurv}. 
The variations of the Hamiltonian that follow from the Lagrangian \eqref{lagqed} are (see appendix \ref{detailsqed} for further details)
\beq
\label{photonaa}
\partial_{\g^2} H = \frac{1}{\g^4} \int d^3x \, (\vec E^2 - \vec B^2) ~, ~~ 
\partial_\theta H = \frac{1}{8\pi^2} \int d^3x \, \vec E\cdot \vec B~,
\eeq
where $\vec E$ and $\vec B$ are the electric and magnetic fields.
Equivalently, in complexified notation
\beq
\label{qedhvar}
\partial_{\tau} H =-\frac{i}{64 \pi} F_+^2~, ~~ \partial_{\overline{\tau}}H = \frac{i}{64\pi}F_-^2~.
\eeq

In this example the Hilbert space is a freely generated Fock space. Consequently, the Berry phase for a multi-photon state is simply the sum of the Berry phases of the individual photons.
As a result, it is sufficient to compute the Berry phase of a single photon, which is labeled by its Kaluza-Klein momentum $\vec{p}$ and its helicity $\epsilon = \pm$. The quantity we want to compute is
\beq
\label{photonab}
\bF^{(\vec{p},\vec{p'},\epsilon,\epsilon')}_{\tau \otau} = \sum_{E_m \neq E_p} \frac{\langle \vec{p}, \epsilon| \partial_{\tau} H | m\rangle \langle m| \partial_{\overline{\tau}} H |\vec{p}{\, '},\epsilon'\rangle - (\tau \leftrightarrow \overline{\tau})}{(E_p - E_m)^2}~.
\eeq

First, we expand the fields in creation and annihilation operators for (on-shell) photons. In particular, the variations of the Hamiltonian \eqref{qedhvar} are quadratic in the creation/annihilation operators. Then, the intermediate states $|m\rangle$ that can contribute in the sum above are only the states that possess one or three photons. 
After the necessary algebra and the explicit evaluation of the sum over states, which is further described in appendix \ref{detailsqed}, one is led to the result
\beq
\label{photonac}
\bF^{(\vec{p},\vec{p'},\epsilon,\epsilon')}_{\tau \otau}  =\frac{\epsilon}{8} \frac{1}{({\rm Im}\tau)^2}
\delta_{\epsilon,\epsilon'}\delta_{\vec{p},\vec{p'}}~.
\eeq
This expression is independent of the momentum $\vec p$, but depends on $\epsilon=\pm 1$, which continues to label the helicity of the photon. Only states with $\vec{p}=\vec{p'}$ and $\epsilon=\epsilon'$ produce a non-vanishing curvature component. Interestingly, the $\tau$-dependent factor is proportional to the Riemann tensor of the parameter space that follows from the metric \eqref{hypmet}.
We will soon see that this relation with the Riemann curvature of the parameter space is true in other examples too.

For a general multi-photon state $|{\mathfrak n}_+,{\mathfrak n}_-\rangle$ that
contains ${\mathfrak n}_+$ (${\mathfrak n}_-$) photons of positive (negative) helicity of arbitrary momentum, the Berry curvature 
follows immediately as
\beq
\label{photonad}
\bF^{({\mathfrak n}_+,{\mathfrak n}_-)}_{\tau \overline{\tau}} = \frac{1}{8} ({\mathfrak n}_+ - {\mathfrak n}_-) \frac{1}{({\rm Im}\tau)^2}~.
\eeq

\subsection[A global effect]{A global effect\footnote{We would like to thank D.Tong for discussions on this topic.}}

As an interesting example consider a closed loop in parameter space which runs from $\theta=0$ to $\theta =2\pi$ at a fixed value of $\g=\g_0$. Using the formulae we derived in the previous subsections we find that a state with ${\mathfrak n}_+$ (${\mathfrak n}_-$) photons of positive (negative) helicity will pick up a phase $e^{i\phi}$, where 
\beq
\label{berryfortheta}
\phi = \int_{\cal D} d\tau d\overline{\tau} \,\,\bF^{({\mathfrak n}_+,{\mathfrak n}_-)}_{\tau \overline{\tau}}~.
\eeq
${\cal D}$ is the domain $0\leq \theta \leq 2\pi$ and $g\leq g_0$ (or ${\rm Im\tau}\geq {\rm Im\tau_0}$). In \eqref{berryfortheta} we used the Stokes theorem to convert the $\theta$-integral over the connection into an integral over the curvature in the interior of the loop.\footnote{A more careful analysis shows that there is no $\delta$-function-like contribution to the curvature from the point ${\rm Im}\tau = \infty$.}
Using \eqref{photonad} we find
\beq
\label{globalberry}
\phi = \frac{1}{16\pi} \g_0^2 ({\mathfrak n}_+ -  {\mathfrak n}_-)~.
\eeq
This relation predicts that a photon state with net helicity will exhibit an overall geometric phase shift \eqref{globalberry} as light travels through a material where $\theta$ varies slowly from $0$ to $2\pi$ at fixed $\g=\g_0$.
As we describe in the next subsection \ref{polrot} for a linearly polarized photon this will have the effect of a rotation of the polarization plane.
Slow variation of $\theta$ (and $\g$ in general) refers to the conditions required by the adiabatic theorem
\beq
\label{adiabaticConditions}
\left | \langle m | \frac{d H}{d t} | k \rangle \right| \ll \frac{|E_k - E_m|}{\Delta T_{km}}~,
\eeq
where $\Delta T_{km}$ is the characteristic time of transition between the states $k,m$. 

We note in passing that the global loop on the parameter space that we consider here is the one that would lead to the relabeling of dyon states via the Witten effect \cite{Witten:1979ey}, though of course our considerations apply only to photon states.

\subsection{Rotation of the polarization plane}
\label{polrot}

In the previous section we found that under adiabatic cyclic variations of $\g,\theta$ photons pick up a phase depending on their helicity. In this subsection we consider an interpretation of this effect in a basis of linearly polarized photons. 

For concreteness consider a linearly polarized photon with momentum $p_z$ along the positive $z$-axis. The polarization of the photon is characterized by a unit vector on the $xy$ plane. With an appropriate choice of conventions, a photon with linear polarization along the $x$-axis is described as a superposition of circularly polarized photons
\beq
|p_z, \hat{x}\rangle = \frac{1}{\sqrt{2}} \left[ |p_z, +\rangle +|p_z, -\rangle    \right]~.
\eeq
A cyclic variation in parameter space will lead to a phase $e^{i \phi}$ for the circularly polarized photons (see eq. \eqref{globalberry}) that transforms this state into
\beq
|p_z, \hat{x}\rangle ~ \longrightarrow ~ |p_z, \hat{\phi}\rangle = 
 \frac{1}{\sqrt{2}} \left[ e^{i \phi} |p_z, +\rangle + e^{-i \phi}|p_z, -\rangle    \right]~.
\eeq
This is a state of linear polarization along an axis $\hat{\phi}$, which is rotated clockwise on the $xy$ plane relative to the $x$-axis. Notice that if we flip the sign of the momentum $p_z$ and consider the same path in parameter space, then the polarization vector will be rotated counter-clockwise on the $xy$ plane.

\subsection{Potential realization of the $U(1)$ Berry phase}

We point out in passing a notable appearance of $\theta$ in the context of magneto-electric properties of solids, where $\theta$ affects the so-called magneto-electric polarizability coefficients (see e.g.\ \cite{Essin:2008rq,Qi:2008ew})
\beq
\label{lagqeda}
\alpha_{ij} = \frac{\d M_j}{\d E_i} \bigg |_{{\vec B}=0} = \frac{\d P_j}{\d B_i} \bigg |_{{\vec E}=0}~.
\eeq
The trace part of $\alpha_{ij}$ is proportional to $\theta e^2/\hbar$. Interestingly, $\theta$ arises here as an integral in momentum space
\beq
\label{lagqedb}
\theta =- \frac{1}{2\pi} \int d^3k \, \varepsilon_{ijk} \Tr \left[ \AA_i \frac{\d \AA_k}{\d k_j} - \frac{2}{3} \AA_i \AA_j \AA_k \right]
\eeq
of a Chern-Simons integrand expressed in terms of another Berry connection, the connection $(\AA_j)_{\mu\nu} = \langle u_\mu | \frac{\d}{\d k_j} | u_\nu \rangle$ of the cell Bloch states $|u_\mu\rangle$ in the occupied bands $\mu$. The trace is accordingly performed over the occupied bands. In $T$-invariant materials the angle $\theta$ takes only two possible values, $\theta=0,\pi$. Topological insulators are characterized by $\theta=\pi$.
When time-reversal is broken, $\theta$ can be varied continuously. Its value depends on the band structure of the material according to \eqref{lagqedb}. We refer the reader to refs.\ \cite{LiWangZhang,Xuetal} for a discussion of setups with varying $\theta$.

\section[4d ${\cal N}=2$ QFTs on the Coulomb branch]{4d ${\cal N}=2$ QFTs on the Coulomb branch\footnote{This section was developed after discussions with C.Vafa, who prompted us to investigate the Berry phase on the Coulomb branch of ${\cal N}=2$ theories.}}
\label{coulomb}

$\NN=2$ QFTs are generically endowed with continuous spaces of vacua (moduli spaces), which are parametrized by the vacuum expectation value (vev) of appropriate operators. For example, $\NN=2$ supersymmetric gauge theories typically possess a Coulomb branch of vacua parametrized by the vev of gauge-invariant combinations of the adjoint complex scalar field(s) in the $\NN=2$ vector multiplet(s). The low-energy effective field theory of these vacua is determined non-perturbatively up to second order in derivatives by Seiberg-Witten theory \cite{Seiberg:1994rs}.

In this section we are concerned with the Berry phase associated to the variation of these vevs on the moduli space. We focus on vacua in the Coulomb branch.

\subsection{Coulomb branch as the parameter space of effective field theory}

The scalar vevs that parametrize the position on the Coulomb branch control the effective couplings of the low energy theory. Hence, from the point of view of the low energy theory, these vevs can be thought of as parameters in an effective Hamiltonian, which will lead to a Berry phase when varied adiabatically.

In order to make this computation precise, it is necessary to deal with a few important subtleties. An honest moduli space of vacua, where the scalars have well defined vevs, arises only in the limit where the volume of space is infinite.
This creates a tension with the IR issues that arise in the computation of the Berry phase due to infinite volume (related to the normalization of states etc.), as we pointed out above. The strategy that requires placing the theory in finite volume, e.g.\ on a torus $\IT^3$, will not work automatically in this case. 

Our attitude in the following subsections will be the following. The theory will be placed on finite, but large volume, where states characterized by fixed scalar vevs are almost ground states whose corresponding wavefunctions spread out slowly by a rate suppressed by the large volume. We will compute the Berry phase to leading order in an approximation where the wavefunction spreading is ignored.

\subsection{Pure ${\cal N}=2$ SU(2) gauge theory}

The $\NN=2$ SYM theory with $SU(2)$ gauge group is characterized by a 1-dimensional Coulomb branch parametrized by $u = \langle {\rm Tr} \Phi^2\rangle$. The Coulomb branch has two singularities $ u = \pm \Lambda$, where extra massless states appear \cite{Seiberg:1994rs}. Away from these singularities the IR theory is an ${\cal N}=2$ $U(1)$ gauge theory, which consists of a massless scalar $a$, the gauge field $A_\mu$ and a set of corresponding fermions. The IR theory is characterized by an effective gauge coupling and theta angle combined in the complex coupling $\tau = \frac{\theta}{2\pi} + i \frac{4\pi}{g^2}$. 
The effective coupling is determined by the low energy prepotential ${\cal F}(a)$ as
\beq
{\cal \tau}(a) = \frac{\partial^2 {\cal F}}{\partial a^2}~.
\eeq
In the approximation discussed in the previous subsection, we can think of the coordinate on the Coulomb branch $a$ as an ``effective parameter'' of the IR theory. We imagine that we vary $a$ adiabatically and we are interested in the resulting Berry phase for various states.
It is easy to see that the Berry curvature for an IR photon of positive helicity is characterized by a 2-form on the Coulomb branch with components
\beq
\bF_{a\overline{a}} = \frac{\partial\tau}{\partial a}  \frac{\partial\overline{\tau}}{\partial \overline{a}} F_{\tau \overline{\tau}} =   \frac{1}{8}  \frac{\partial\tau}{\partial a}  \frac{\partial\overline{\tau}}{\partial \overline{a}} 
\frac{1}{({\rm Im}\tau)^2}
~,
\eeq
where $F_{\tau \overline{\tau}}$ above was evaluated using \eqref{photonac}. Now, remember that the metric on the Coulomb branch is
\beq
g_{a \overline{a}} = {\rm Im} \frac{\partial^2 {\cal F}}{\partial a^2} = {{\rm Im} \tau}~,
\eeq
and notice that
\beq
\frac{\partial \tau}{\partial a} = \frac{\partial}{\partial a} (\tau - \overline{\tau}) = 2i \frac{\partial}{\partial a} {\rm Im} \tau = 2i \partial_a g_{a \overline{a}}~.
\eeq
As a result, the above formula for the Berry curvature can be written as
\beq
\bF_{a\overline{a}} =  \frac{1}{2}  g^{\overline{a}a} g^{ \overline{a}a} \partial_a g_{a\overline{a}} \partial_{\overline{a}}g_{a \overline{a}}~,
\eeq
where we recognize the expression on the right hand side as the Riemann curvature on the Coulomb branch.

Hence, we find that the Berry curvature of a photon of positive helicity is proportional to the Riemann tensor on the Coulomb branch
\beq
\bF_{a \overline{a}} = \frac{1}{2}  R^{a}_{a a \overline{a}}~.
\eeq
Obviously a negative helicity photon has the opposite Berry phase.

In this section we computed the Berry phase of IR photons on the Coulomb branch of ${\cal N}=2$ theories. The IR spectrum of the theory also contains massless fermions and scalars, belonging to the same ${\cal N}=2$ vector multiplet. Supersymmetry requires that the Berry phase of all states in the same supermultiplet should be related. It would be interesting to directly compute the Berry phase of the low energy scalars and fermions.

\subsection{Generalization to higher rank Coulomb branch}
\label{sec:n2coulomb}

Next, let us consider a 4d ${\cal N}=2$ theory with a Coulomb branch of rank $r$. The scalar fields are $a^i$, with $i=1,...,r$. The IR $U(1)^r$ couplings are characterized by the matrix
\beq
\tau_{ij} = \frac{\partial^2 {\cal F}}{\partial a^i \partial a^j}~,
\eeq
which parametrizes the theta angles and gauge couplings of the IR photons
\beq
\label{photonae}
\LL = -\frac{1}{4 g^2_{ij}} F^i_{\mu\nu} F^{j, \mu\nu} + \frac{\theta_{ij}}{32 \pi^2} F_{\mu\nu}^i \widetilde{F}^{j,\mu\nu} ~,
\eeq
or in complex notation
\beq
\label{photonaf}
\LL = \frac{i}{64\pi} \tau_{ij} F_+^i F_+^j  - \frac{i}{64 \pi} \overline{\tau}_{ij} F_-^i F_-^j~.
\eeq
The matrix $\tau_{ij}$ is a symmetric $r\times r$ matrix with positive imaginary part. The metric on the Coulomb branch is
\beq
g_{i\overline{j}} = {\rm Im} \tau_{ij}~.
\eeq

An IR photon in this theory will be labeled as $|\vec{p},\epsilon; i\rangle$, where the last label refers to each $U(1)$ gauge group individually. Following the same steps as before, we find that photons in the infrared are characterized by a {\it non-abelian} Berry phase, whose curvature is\footnote{Here $k,\overline{l}$ denote the tangent directions along the moduli space, while $i,j$ are the indices of the photons whose (non-abelian) Berry phase we are computing.}
\beq
\label{berrysw}
(\bF_{k\overline{l}})_j^i=  \frac{1}{2} R^{i}_{jk \overline{l}}~.
\eeq
Once again this is proportional to the Riemann tensor on the Coulomb branch. We conclude that the curvature of the Seiberg-Witten metric characterizes the Berry phase that a low-frequency photon receives under an adiabatic loop in the Coulomb branch.

\section{Massive ${\cal N}=1$ theories on $\IR \times \IT^3$}
\label{massive}

Our next focus is the Berry formula for supersymmetric ground states in 4d $\NN=1$ massive theories.
The deformations of interest preserve the $\NN=1$ supersymmetry and are induced on the level of a Lagrangian by $F$-terms of the form
\beq
\label{N1aa}
\delta \LL = \lambda^i \, Q^2 \cdot \varphi_i + \bar\lambda^i \, \bar Q^2 \cdot \varphi_i~,
\eeq
where $Q_\alpha$, $\bar Q_{\dot \alpha}$ are the four real supercharges of the theory, $Q^2$ etc.\ denote the nested action of the supercharges, and $\varphi_i$ are chiral primary operators. The deformation is classically marginal or relevant when the UV scaling dimension of the operators $\varphi_i$ is less than or equal to 3.

A particularly important example to keep in mind is $\NN=1$ SYM theory with gauge group $SU(N)$. In this case we may consider deformations by the super-Yang-Mills interaction
\beq
\label{N1ab}
\delta \LL = \frac{1}{32\pi} {\rm Im} \left[ \tau \int d^2 \vartheta\, \Tr \left( W^\alpha W_\alpha \right) \right]~,
\eeq
where $\tau= \frac{\theta}{2\pi} + \frac{4\pi}{g_{YM}^2}$ is as before the complexified Yang-Mills coupling and $W_\alpha$ the chiral superfield whose bottom component is the gaugino $\chi_\alpha$. The chiral primary operator $\varphi$ that implements \eqref{N1aa} is the gaugino composite operator
\be
\label{N1ac}
\varphi = \Tr \left[ \chi^\alpha \chi_\alpha \right]~.
\eeq
As is well known, the $\NN=1$ SYM theory is asymptotically free and the interaction \eqref{N1ab} is quantum mechanically relevant. Under renormalization group flow the theory develops a dynamically generated strong coupling scale 
\be
\label{N1ad}
\Lambda = \mu\, e^{-\frac{8\pi^2}{3N g_{YM}^2(\mu)}}~,
\ee
where $\mu$ is a reference energy scale. Hence, the deformations \eqref{N1ab} of the theory can be viewed as deformations of the strong coupling scale $\Lambda$, or on $\IR\times \IT^3$ deformations of the dimensionless quantity $R \Lambda$, where $R$ is the radius of $\IT^3$.

On $\IR \times \IT^3$ the Hamiltonian $H$ is 
\beq
H = P_0 = i\partial_t ~.
\eeq
The states whose Berry phase we are interested in computing in this section are ground states (namely zero energy eigenstates) of this Hamiltonian.
The relative (i.e.\ bosonic$-$fermionic) number of these states is counted by the Witten index \cite{Witten:1982df}.

In the example of $\NN=1$ SYM theory, recall that on $\IR^4$ the theory exhibits $N$ discrete vacua labeled by the value of the chiral condensate
\be
\label{N1ae}
\langle \varphi \rangle = \Lambda^3 \, e^{\frac{2\pi i n}{N}}~, ~~ n =0,1,2,\ldots, N-1~.
\eeq
Consequently, for this theory we are interested in the Berry phase associated with the vacua \eqref{N1ae} under adiabatic changes of $\Lambda$.

Returning to the general situation, let us denote the ground states as $| I\rangle$ and proceed as follows. To keep the notation short, the integrals $\oint$ will denote integrals on $\IT^3$. $\HH_0$ will denote the Hilbert subspace of the ground states. 

For the holomorphic-holomorphic components of the curvature we have
\begin{align}
\left( \bF_{kl} \right)_{IJ} =  \sum_{n,m\not \in \HH_0} 
\frac{1}{E_n^2}
\langle J | \left( Q^2 \cdot \oint \varphi_k \right) |m\rangle g^{mn}
\langle n | Q^2  \cdot \oint \varphi_l \, |I \rangle 
-(k\leftrightarrow l)~.
\end{align}
Here, but also in the following sections, it will be convenient to define the auxiliary quantity
\begin{align}
\label{N1ag}
&\left( \tilde \bF_{kl} \right)_{IJ}=
\langle J | \left( Q^2 \cdot \oint \varphi_k \right) (H-x)^{-2} \left( Q^2 \cdot \oint \varphi_l \right) |I \rangle 
-(k \leftrightarrow l)
\cr
&=  \sum_{n,m\not \in \HH_0} 
\frac{1}{(E_n - x)^2}
\langle J | \left( Q^2 \cdot \oint \varphi_k \right) |m \rangle g^{mn} 
\langle n | Q^2  \cdot \oint \varphi_l \, |I \rangle 
-(k\leftrightarrow l)
\cr
&+ \sum_{n,m\in \HH_0} 
\frac{1}{(E_n - x)^2}
\langle J | \left( Q^2 \cdot \oint \varphi_k \right) |m\rangle g^{mn}
\langle n | Q^2  \cdot \oint \varphi_l \, |I \rangle 
-(k\leftrightarrow l)~,
\end{align}
where $x$ is an auxiliary free real parameter. Since the chiral supercharges $Q$ annihilate the bra and ket ground states the last line does not contribute and we conclude that
\beq
\left( \bF_{kl} \right)_{IJ} = \lim_{x \to 0} \left( \tilde \bF_{kl} \right)_{I\bar J}~.
\eeq

Now, we notice that since
\beq
\left[ Q_\alpha, P_\mu \right] =0~,
\eeq
the following commutation relation holds
\be
(H-x)^{-2} Q^2 = Q^2 (H-x)^{-2}~.
\ee
Consequently, we can move $Q^2$ on the right integral in \eqref{N1ag} (see first line) across $(H-x)^{-2}$ towards the left. On the left it annihilates everything yielding
\be
\label{N1ba}
\left( \bF_{kl} \right)_{IJ} =0~.
\ee
In a similar fashion we can show that
\be
\label{N1bb}
\left( \bF_{\bar k \bar l} \right)_{IJ} =0~.
\ee

The remaining mixed components of the curvature, $(\bF_{k\bar l})_{IJ}$, are more interesting.
Repeating the above steps we first define
\be
\left( \bF_{k\bar l} \right)_{IJ} = \lim_{x \to 0} \left( \tilde \bF_{k\bar l} \right)_{IJ}~.
\ee
Then, obvious manipulations with the supercharges yield
\begin{align}
\left( \tilde \bF_{k\bar l} \right)_{IJ}
&=  \langle J | \left( Q^2 \cdot \oint \varphi_k \right) (H-x)^{-2} 
\left( \bar Q^2 \cdot \oint \bar \varphi_{\bar l} \right) |I \rangle 
-(k \leftrightarrow \bar l)
\cr
&= \langle J | \left[ \bar Q^2 \cdot \left( Q^2 \cdot \oint \varphi_k \right)\right] (H-x)^2 \oint \bar \varphi_{\bar l}  
|I \rangle 
-(k \leftrightarrow \bar l)
\cr
&= \kappa 
\langle J | \oint \nabla^2 \varphi_k \, (H-x)^2 \oint \bar \varphi_{\bar l}  |I \rangle 
-(k \leftrightarrow \bar l)~,
\end{align}
where we used 
\beq
\bar Q^2 \cdot Q^2 = \kappa \nabla^2~.
\eeq
$\kappa$ is a numerical constant whose precise value depends on the normalization conventions for the supercharges.
In what follows we will set this constant to 1. Assuming we can ignore terms with total space derivatives
\be
\oint \d^i \d_i \varphi_k =0~, ~~ i=1,2,3
\ee
inside the correlation functions, we finally obtain
\begin{align}
\left( \tilde \bF_{k\bar l} \right)_{IJ}
= - \langle J | \oint \varphi_k \, H^2 (H-x)^{-2} \oint \bar \varphi_{\bar l}  |I \rangle 
-(k \leftrightarrow \bar l)~.
\end{align}
After the limit $x\to 0$ we find
\bea
\left( \bF_{k\bar l} \right)_{IJ}&=&- \sum_{n,m\not\in \HH_0}
\langle J | \oint \varphi_k | m\rangle g^{mn}  \langle n | \oint \bar\varphi_{\bar l} | I \rangle 
-(k \leftrightarrow \bar l)
\cr
&=&- 
\bigg\{
\langle J | \left[ \oint \varphi_k, \oint \bar \varphi_{\bar l} \right] | I\rangle 
- \sum_{M,N \in \HH_0} 
\left[  \langle J | \oint \varphi_k | M\rangle g^{MN} \langle N | \oint \bar \varphi_{\bar l} | I \rangle 
-(k \leftrightarrow \bar l)  \right]
\bigg\}
~.\nonumber\\
\eea

The first term on the r.h.s.\ of this equation is a contact term, while the sum in the second term is expressed in terms of the transition amplitudes $g_{MN}=\langle M | N\rangle$ and the vevs of the (anti)chiral primaries
\be
\label{N1cb}
C_{kMN}=\langle N | \oint \varphi_k | M\rangle ~,~~
C^*_{\bar l M N} = \langle N | \oint \varphi_{\bar l} | M\rangle~.
\ee
As a consequence, the Berry curvature assumes the very simple form 
\be
\label{N1cc}
\left( \bF_{k\bar l} \right)_{IJ}
=-   \left[ C_k, \bar C_{\bar l} \right]_{I\bar J} -
\langle J | \left[ \oint \varphi_k, \oint \bar \varphi_{\bar l} \right] | I\rangle~,
\ee
where $\left[ C_k, \bar C_{\bar l} \right]_{I\bar J} = C_{kI}^P g_{P\bar Q} C^{*\bar Q}_{\bar l \bar J} - g_{P\bar J}C_{kV}^P g^{V \bar U} C^{*\bar N}_{\bar U \bar l} g_{I\bar N}$. This equation exhibits the same structure as the \tts equations \cite{Cecotti:1991me,Papadodimas:2009eu}.
It would be interesting to evaluate explicitly both terms on the r.h.s.\ of equation \eqref{N1cc}, and understand the corresponding physics in more detail in specific examples, such as the $\NN=1$ SYM theory. We hope to return to this problem in a different publication.

\section{Berry phase in $2d$ $\NN=(2,2)$ SCFTs}
\label{2d}

In this section (and the next) we slightly change gears and proceed with an explicit evaluation of the Berry curvature formula \eqref{revbb} in (super)conformal field theories. This provides another general example of QFTs that exhibit rich Berry-like properties.
We consider the CFT in radial quantization (equivalently, the CFT is formulated on $\IR\times S^{d-1}$) and implement the operator-state correspondence. This allows us to establish a natural relation between the quantum mechanics Berry phase and previous results on operator mixing in conformal perturbation theory. We will discuss the general features of this relation for arbitrary CFTs in section \ref{specVScft}. 

We begin with the evaluation of the Berry phase of chiral primary states in the NSNS sector of 2d $\NN=(2,2)$ SCFTs. The Berry curvature in the RR sector was first computed by Cecotti and Vafa many years ago in \cite{Cecotti:1991me}. A related formula was derived for the NS sector within conformal perturbation theory in Ref.\ \cite{deBoer:2008ss} by evaluating the 4-point function formula \eqref{specad}. We will now show that the quantum mechanics perspective \eqref{revbb} leads to the same result.

The chiral states, whose Berry phase we want to compute, are characterized by the conditions\footnote{Here we focus for definiteness on the left-moving sector. The same relations are also obeyed on the right-moving sector.}
\beq
\label{2daa}
\QQ^\pm |I\rangle =0~, ~~ \SS^- | I \rangle =0~.
\eeq
The precise definitions of the supercharges in the $\NN=(2,2)$ superconformal algebra and their (anti)commutation relations are summarized in appendix \eqref{conv2dN2}. The chiral bra states, which are denoted here as $\langle \bar I |$, obey
\beq
\label{2dab}
\langle \bar I | \QQ^\pm =0~, ~~ \langle \bar I | \SS^+ =0~.
\eeq

We want to consider infinitesimal deformations of the Hamiltonian by exactly marginal $F$-term deformations with vanishing $U(1)_R$ charge and vanishing energy. This requirement implies the general form of Hamiltonian deformations
\beq
\label{2dac}
\delta H = \frac{\delta \lambda^i}{2\pi} \oint \QQ^- \bar\QQ^- \cdot \varphi_i 
+\frac{\delta \bar\lambda^{\bar i}}{2\pi} \oint \QQ^+\bar\QQ^+ \cdot \bar \varphi_{\bar i}~.
\eeq
The coordinates $(\lambda^i,\bar\lambda^{\bar i})$ provide a local parametrization of the superconformal manifold of the 2d SCFT.  
The action of the supercharges $\QQ^\pm$, $\bar \QQ^\pm$ on the operators $\varphi_i$, $\bar\varphi_{\bar i}$ denotes the appropriate nested (anti-)commutator. The operators $\varphi_i$ are rotated versions (see eq.\ \eqref{convad} in appendix \ref{conventions}) of chiral primary operators with equal left/right scaling dimension $h_L=h_R=1$. Similarly, $\bar\varphi_{\bar i}$ are rotated versions of anti-chiral primary operators with $h_L=h_R=1$. 

With these specifications we can proceed to determine the quantity of interest
\beq
\label{2dad}
\left( \bF_{\mu\nu} \right)_{I\bar J} =  \sum_{n \not\in \HH_I} \sum_{a,\bar b\in \HH_n}
\frac{1}{(E_I-E_n)^2} \langle \bar J | \d_\mu H |n,a \rangle
g_{(n)}^{\bar b a}
\langle n, \bar b | \d_\nu H | I \rangle - (\mu \leftrightarrow \nu)~.
\eeq
The indices $\mu,\nu$, which parametrize different directions in the parameter space $(\lambda^i,\bar \lambda^{\bar i})$, can be either holomorphic or anti-holomorphic. 
We discuss each of the possible cases separately.

When both $\mu$ and $\nu$ are holomorphic we obtain (after using \eqref{2dac})
\beq
\label{2dae}
\left( \bF_{kl} \right)_{I\bar J} = \frac{1}{(2\pi)^2} \sum_{n\not \in \HH_I}  \sum_{a,\bar b\in \HH_n}
\frac{1}{(E_I-E_n)^2} 
\langle \bar J | \oint \QQ^- \bar \QQ^-\cdot \varphi_k |n,a \rangle
g_{(n)}^{\bar b a}
\langle n,\bar b | \oint \QQ^- \bar \QQ^-\cdot \varphi_l | I \rangle - (k \leftrightarrow l)~.
\eeq
Similar to the previous section, it is convenient to introduce an auxiliary parameter $x$ and define the quantity
\beq
\label{2daf}
\left( \tilde \bF_{kl} \right)_{I\bar J} = \frac{1}{(2\pi)^2}  
\langle \bar J | \left( \oint \QQ^- \bar \QQ^-\cdot \varphi_k \right)
\left ( H-\frac{1}{2} R - x\right)^{-2}
\left( \oint \QQ^- \bar \QQ^-\cdot \varphi_l \right) | I \rangle - (k \leftrightarrow l)~.
\eeq
$(\bF_{kl})_{I\bar J}$ can be easily recovered from $\left( \tilde \bF_{kl} \right)_{I\bar J}$ by taking the limit $x\to 0$
\beq
\label{2dag}
\left( \bF_{kl} \right)_{I\bar J} = \lim_{x\to 0} \left( \tilde \bF_{kl} \right)_{I\bar J}~.
\eeq
Employing the commutation relation
\beq
\label{2dai}
\left[ H -\frac{1}{2} R, \QQ^- \right]=0
\eeq
and the fact that $\QQ^-$ annihilates both external states we deduce immediately that $\left( \tilde \bF_{kl}\right)_{I\bar J}=0$. These observations allow us to obtain trivially the identities
\beq
\label{2daj}
\left( \bF_{kl}\right)_{I\bar J}=0~,~~
\left( \bF_{\bar k\bar l}\right)_{I\bar J}=0~.
\eeq
The second identity follows in exactly the same fashion as the first.

The case of mixed components is more interesting:
\beq
\label{2dak}
\left( \bF_{k\bar l} \right)_{I\bar J} = \frac{1}{(2\pi)^2} \sum_{n\not \in \HH_I}  \sum_{a,\bar b\in \HH_n}
\frac{1}{(E_I-E_n)^2} 
\langle \bar J | \oint \QQ^- \bar \QQ^-\cdot \varphi_k |n,a \rangle
g_{(n)}^{\bar b a}
\langle n,\bar b | \oint \QQ^+ \bar \QQ^+\cdot \bar \varphi_{\bar l} | I \rangle - (k \leftrightarrow \bar l)~.
\eeq
Again, we express this quantity as the limit
\beq
\label{2dal}
\left( \bF_{k\bar l} \right)_{I \bar J} = \lim_{x\to 0} \left( \tilde \bF_{k \bar l} \right)_{I\bar J}~,
\eeq
with
\beq
\label{2dam}
\left( \tilde \bF_{k\bar l} \right)_{I\bar J} = \frac{1}{(2\pi)^2}  
\langle \bar J | \left( \oint \QQ^- \bar \QQ^-\cdot \varphi_k \right)
\left ( H-\frac{1}{2} R - x\right)^{-2}
\left( \oint \QQ^+ \bar \QQ^+\cdot \bar \varphi_{\bar l} \right) | I \rangle - (k \leftrightarrow \bar l)~.
\eeq
Then we can use the commutation
\beq
\label{2dan}
\left[ H -\frac{1}{2}R ,\QQ^+ \right] =0
\eeq
(and its right-moving version) to move the supercharges $\QQ^+$, $\bar \QQ^+$ over $(H-\frac{1}{2}R-x)^{-2}$ to the left. On the left both of these supercharges annihilate the bra $\langle \bar J |$ and since they commutate with the chiral primary operator $\varphi_k$ we deduce trivially the expression
\beq
\label{2dao}
\left( \tilde \bF_{k\bar l} \right)_{I\bar J} = \frac{1}{(2\pi)^2}  
\langle \bar J | \left( \oint \{ \QQ^-,\QQ^+\} \{ \bar \QQ^-, \bar \QQ^+ \} \cdot \varphi_k \right)
\left ( H-\frac{1}{2} R - x\right)^{-2}
\oint \bar \varphi_{\bar l} \, | I \rangle - (k \leftrightarrow \bar l)~.
\eeq
Implementing the first identity of eq.\ \eqref{conv2dN2acb} we further obtain
\beq
\label{2dap}
\left( \tilde \bF_{k\bar l} \right)_{I\bar J} = \frac{4}{(2\pi)^2}  
\langle \bar J | \left[\left( \LL_0 -\frac{1}{2}\JJ_0 \right) \left( \bar \LL_0 -\frac{1}{2}\bar\JJ_0 \right), \oint \varphi_k \right] 
\left ( H-\frac{1}{2} R - x\right)^{-2}
\oint \bar \varphi_{\bar l} \, | I \rangle - (k \leftrightarrow \bar l)~,
\eeq
where $\LL_0, \JJ_0$ etc.\ are modes of Virasoro and $U(1)_R$ generators (see appendix \ref{conv2dN2} for further details on the notation).
Notice, however, that since the chiral insertion $\phi_k$ is spinless with equal left/right $U(1)_R$ charges one can easily deduce from the identity
\beq
\label{2daq}
\left( \LL_0 -\frac{1}{2}\JJ_0 \right) \left( \bar\LL_0 -\frac{1}{2} \bar\JJ_0 \right) =
\frac{1}{4} \left( \HH_+^2 -\HH_-^2 \right)~,
\eeq
\beq
\label{2dar}
\HH_+ = H-\frac{1}{2}R~, ~~ 
\HH_- = \LL_0 -\bar\LL_0 -\frac{1}{2}(\JJ_0 -\bar \JJ_0)~,
\eeq
that 
\bea
\label{2das}
\left( \tilde \bF_{k\bar l} \right)_{I\bar J} &=& \frac{1}{(2\pi)^2}  
\langle \bar J | \left[\left( H -\frac{1}{2}R \right)^2, \oint \varphi_k \right] 
\left ( H-\frac{1}{2} R - x\right)^{-2}
\oint \bar \varphi_{\bar l} \, | I \rangle - (k \leftrightarrow \bar l)
\nonumber\\
&=& -\frac{1}{(2\pi)^2}  
\langle \bar J | \oint \varphi_k  
\left( H -\frac{1}{2}R \right)^2 
\left ( H-\frac{1}{2} R - x\right)^{-2}
\oint \bar \varphi_{\bar l} \, | I \rangle - (k \leftrightarrow \bar l)~.
\eea
As a result, by taking the limit $x\to 0$ we find
\bea
\label{2dat}
\left( \bF_{k\bar l} \right)_{I\bar J} &=&- \frac{1}{(2\pi)^2} 
\sum_{n,m\not \in \HH_{chiral}} \langle \bar J | \oint \varphi_k |n \rangle g^{\bar m n} 
\langle \bar m | \oint \bar\varphi_{\bar l} |I\rangle
-(k \leftrightarrow \bar l)
\\
&=& -\frac{1}{(2\pi)^2} \bigg\{
\langle  \bar J | \left[ \oint \varphi_k , \oint \bar\varphi_{\bar l} \right] | I\rangle 
\cr
&&
- \sum_{M, P \in \HH_{chiral}}  \left[
\langle \bar J | \oint \varphi_k | M \rangle g^{\bar P M} \langle \bar P | \oint \bar\varphi_{\bar l} | I\rangle 
- \langle \bar J | \oint \bar \varphi_{\bar l} | M \rangle g^{\bar P M} \langle \bar P | \oint \varphi_k | I\rangle  \right]
\bigg\}~.\nonumber
\eea
$\HH_{chiral}$ refers to the Hilbert subspace of chiral primary states. Clearly, only a finite number of chiral primary states contributes to the last two terms of the above expression, those that saturate the $U(1)_R$ charge conservation equations $R_M=R_P$, $R_M + 2 = R_J$ and $R_M-2=R_J$ respectively.

The last line in \eqref{2dat} is immediately recognizable
\bea
\label{2dau}
&&- \frac{1}{(2\pi)^2} 
\sum_{M, P \in \HH_{chiral}}  \left[
\langle \bar J | \oint \varphi_k | M \rangle g^{\bar P M} \langle \bar P | \oint \bar\varphi_{\bar l} | I\rangle 
- \langle \bar J | \oint \bar \varphi_{\bar l} | M \rangle g^{\bar P M} \langle \bar P | \oint \varphi_k | I\rangle  \right]
\cr
&&
\cr
&&=C_{kI}^P g_{P\bar Q} C^{*\bar Q}_{\bar l \bar J} 
- g_{P\bar J}C_{kV}^P g^{V \bar U} C^{*\bar N}_{\bar U \bar l} g_{I\bar N}
= \left[ C_k, \bar C_{\bar l} \right]_{I\bar J}~,
\eea
where $C_{KL}^M$ are the OPE coefficients for chiral primaries
\beq
\label{2dav}
\varphi_K \, | L\rangle = C_{KL}^M \, |M\rangle ~.
\eeq

The remaining term on the r.h.s.\ of \eqref{2dat}, proportional to
\beq
\label{2dba}
\RR = \frac{1}{(2\pi)^2} \langle  \bar J | \left[ \oint \varphi_k , \oint \bar\varphi_{\bar l} \right] | I\rangle ~,
\eeq
is a contact term. Naively it would appear to vanish, but a careful treatment of the short distance singularities that occur when the integrated operators collide shows that the actual contribution is non-vanishing. In appendix \ref{2dtechnical} we show that
\beq
\label{2dbaa}
\RR = - \left( 1 - \frac{3}{c} (q+\bar q) \right) g_{I \bar J} g_{k\bar l}~.
\eeq

Collecting all the contributions we obtain the final result
\beq
\label{2dbu}
\left( \bF_{k\bar l} \right)_{I\bar J}=
- \left[ C_k, \bar C_{\bar l} \right]_{I\bar J}
+ \left( 1 - \frac{3}{c} (q+\bar q) \right) g_{I \bar J} g_{k\bar l}~,
\eeq
which is the same result for the curvature of the conformal manifold connection as the one obtained in superconformal perturbation theory (using eq.\ \eqref{specad}) in \cite{deBoer:2008ss}.

This result is a satisfying re-derivation of the $tt^*$ equations in 2d $\NN=(2,2)$ superconformal manifolds from standard notions in quantum mechanics. Compared to the derivation in superconformal perturbation theory \cite{deBoer:2008ss}, where one needs to make a judicious use of superconformal Ward identities (see section 4.3 in \cite{deBoer:2008ss}), in the above quantum mechanical derivation we arrived at the key formula \eqref{2dat} in a much more straightforward, technically convenient, manner. In the next section, we show that the same is true in four-dimensional SCFTs.

\section{Berry phase in $4d$ $\NN=2$ SCFTs}
\label{4d2}

Our second example in superconformal field theory is the computation of the Berry phase of chiral primary states in 4d $\NN=2$ SCFTs. Like in the 2d theories that we studied in the previous section, the Berry curvature turns out to be related to the curvature that characterizes operator mixing in conformal perturbation theory. The latter is in turn completely determined by the two- and three-point functions of chiral primary operators \cite{Papadodimas:2009eu}. Thanks to recent developments, these correlation functions are now computable in several 4d $\NN=2$ SCFTs \cite{Baggio:2014sna,Baggio:2014ioa,Baggio:2015vxa,Gerchkovitz:2016gxx,Rodriguez-Gomez:2016ijh,Rodriguez-Gomez:2016cem,Baggio:2016skg,Pini:2017ouj}. Therefore, these results can now be interpreted as an exact determination of the Berry curvature for the chiral primary states of these theories.

Our conventions for the 4d $\NN=2$ superconformal algebra follow closely those in \cite{Minwalla:1997ka,Dolan:2002zh}. They are summarized in appendix \ref{conv4dN2}. In these conventions the chiral ket states $| I\rangle$ satisfy by default the relations
\beq
\label{4d2aa}
\QQ^{-\alpha}_i |I\rangle = 0~, ~~
\SS^{-i\dot\alpha} |I\rangle =0~, ~~
\SS^{+i}_\alpha |I\rangle =0~.
\eeq
The index $i=1,2$ is an $SU(2)_R$ index and the indices $\alpha,\dot\alpha=\pm$ are standard spinor indices. For the chiral bra states $\langle \bar I |$ we have similarly
\beq
\label{4d2ab}
\langle \bar I | \QQ^{+i}_\alpha = 0~, ~~
\langle \bar I | \SS^{-i\dot\alpha} =0~, ~~
\langle \bar I | \SS^{+i}_\alpha =0~.
\eeq
The superconformal algebra generators are defined in equations \eqref{SCAconventions}.

In the same conventions the infinitesimal deformations of the Hamiltonian by exactly marginal $\NN=2$ $F$-term deformations involve interactions that have vanishing energy and $U(1)_R$ charge. They have the general form
\beq
\label{4dac}
\delta H = \frac{\delta \lambda^k}{(2\pi)^2} \left( \SS^- \right)^4 \cdot \oint \varphi_k 
+  \frac{\delta \bar \lambda^{\bar l}}{(2\pi)^2} \left( \SS^+ \right)^4 \cdot \oint \bar \varphi_{\bar l}~.
\eeq
In this expression the action of the supercharges $\SS^\pm$ on the operators $\varphi_k$, $\bar\varphi_{\bar l}$ is again via the appropriate nested anti-commutators. The operators $\varphi_k$, $\bar\varphi_{\bar l}$ are rotated versions (see eq.\ \eqref{convad} in appendix \ref{conventions}) of the standard chiral/anti-chiral primary operators with scaling dimension $\Delta=2$. The parameters $(\lambda^i,\bar\lambda^{\bar i})$ parametrize local patches of the 4d $\NN=2$ superconformal manifold.

We proceed to determine the curvature
\beq
\label{4dad}
\left( \bF_{\mu\nu} \right)_{I\bar J} = \sum_{n \not\in \HH_I} \sum_{a,\bar b\in \HH_n}
\frac{1}{(E_I-E_n)^2} \langle \bar J | \d_\mu H |n,a \rangle
g_{(n)}^{\bar b a}
\langle n, \bar b | \d_\nu H | I \rangle - (\mu \leftrightarrow \nu)~.
\eeq
When both indices $\mu$, $\nu$ are holomorphic we can write
\beq
\label{4dae}
\left( \bF_{kl} \right)_{I\bar J} = \lim_{x\to 0} \left( \tilde \bF_{kl} \right)_{I\bar J}~,
\eeq
where
\beq
\label{4daf}
\left( \tilde \bF_{kl} \right)_{I\bar J} = \frac{1}{(2\pi)^4} 
\langle \bar J |  \left( \SS^- \right)^4 \cdot \oint \varphi_k  
\left( H -\frac{1}{2}R-x\right)^{-2}
\left( \SS^- \right)^4 \cdot \oint \varphi_l  |I\rangle - (k\leftrightarrow l)~. 
\eeq
Then, one can use the identity (see appendix \ref{conventions})
\beq
\label{4dag}
\left[ H - \frac{1}{2}R, \SS^\pm \right]=0
\eeq
for $\SS^-$ to move it across $\left( H -\frac{1}{2}R-x\right)^{-2}$ from the left to the right or vice versa. 
At the new position $\SS^-$ annihilates everything and one arrives trivially at the conclusion that
\beq
\label{4dai}
\left( \bF_{kl}\right)_{I\bar J}=0~.
\eeq
Similarly, one shows that $\left( \bF_{\bar k\bar l}\right)_{I\bar J}=0$. This part works in complete analogy to the 2d $\NN=(2,2)$ case described in the previous section.

Important qualitative differences with the 2d computation arise in the case of the mixed components
\beq
\label{4daj}
\left( \bF_{k\bar l} \right)_{I\bar J} = \frac{1}{(2\pi)^4} \sum_{n\not \in \HH_I}  \sum_{a,\bar b\in \HH_n}
\frac{1}{(E_I-E_n)^2} 
\langle \bar J | \left( \SS^-\right)^4 \cdot \oint \varphi_k |n,a \rangle
g_{(n)}^{\bar b a}
\langle n,\bar b | \left( \SS^+\right)^4 \cdot \oint \bar \varphi_{\bar l} | I \rangle - (k \leftrightarrow \bar l)~.
\eeq
As before, these components can be recast as
\beq
\label{4dak}
\left( \bF_{k\bar l} \right)_{I\bar J} = \lim_{x\to 0} \left( \tilde \bF_{k\bar l} \right)_{I\bar J}~,
\eeq
with
\beq
\label{4dal}
\left( \tilde \bF_{k\bar l} \right)_{I\bar J} = \frac{1}{(2\pi)^4} 
\langle \bar J |  \left( \SS^- \right)^4 \cdot \oint \varphi_k  
\left( H -\frac{1}{2}R-x\right)^{-2}
\left( \SS^+ \right)^4 \cdot \oint\bar \varphi_l  |I\rangle - (k\leftrightarrow \bar l)~.
\eeq
Repeating the previous steps we can use \eqref{4dag} to move all four $\SS^+$'s from the right to the left across the operator insertion $\left( H -\frac{1}{2}R-x\right)^{-2}$. On the left the $\SS^+$'s annihilate the external bra and commute with the chiral primary $\varphi_k$.\footnote{The validity of this commutation at every spacetime point follows from the fact that the $\SS^+$'s are related to the supercharges $\bar Q$ by a similarity transformation, see eq.\ \eqref{conv4dN2aaa}. We would not have been able to perform the same manipulation with $\SS^-$ and $\bar\phi_{\bar l}$ by moving the $\SS^-$'s to the right, since the $\SS^-$ supercharges are obtain by similarity from the superconformal partners $\bar S$. The latter do not commute with the anti-chiral fields at all spacetime points.}
Hence, we obtain
\beq
\label{4dam}
\left( \tilde \bF_{k\bar l} \right)_{I\bar J} = \frac{1}{(2\pi)^4} 
\langle \bar J |  \Big[ \left( \SS^- \right)^4 , \left( \SS^+ \right)^4 \Big] \cdot \oint \varphi_k  
\left( H -\frac{1}{2}R-x\right)^{-2}
\oint\bar \varphi_l  |I\rangle - (k\leftrightarrow \bar l)~.
\eeq
The commutator of the supercharges can be determined using the superconformal algebra relations
\beq
\label{4dan}
\left[ \left( \SS^- \right)^4 , \left( \SS^+ \right)^4 \right] = - \left( H -\frac{1}{2} R\right)^4
+4 \left( H -\frac{1}{2}R\right)^2~.
\eeq
Inserting this expression into \eqref{4dam} we find
\beq
\label{4dao}
\left( \tilde \bF_{k\bar l} \right)_{I\bar J} = \frac{1}{(2\pi)^4} 
\langle \bar J | \oint \varphi_k
\left( \left( H -\frac{1}{2} R\right)^4 - 4 \left( H -\frac{1}{2}R\right)^2 \right)
\left( H -\frac{1}{2}R-x\right)^{-2}
\oint\bar \varphi_l  |I\rangle - (k\leftrightarrow \bar l)~.
\eeq

It is instructive to compare this formula with its 2d $\NN=(2,2)$ analog \eqref{2das}. Notice that the 4d formula \eqref{4dao} includes a term that involves the operator $\frac{(H-\frac{1}{2}R)^4}{(H-\frac{1}{2}R-x)^2}$, which does not have an analog in the 2d formula \eqref{2das}.

At this point we can use the fact that
\beq
\label{4dap}
\frac{\left( H-\frac{1}{2}R \right)^n }{\left( H -\frac{1}{2}R - x\right)^2} |I\rangle=0~, ~~ n\in {\mathbb Z}_+
\eeq
and the commutation relations
\begin{subequations}
\bea
\label{4daq}
\left[ H -\frac{1}{2}R, \varphi_{k} \right]&=&\left(\d_\tau -2 \right) \varphi_{k}~,
\\
\left[ H -\frac{1}{2}R, \bar\varphi_{\bar l} \right]&=&\left(\d_\tau +2 \right) \bar\varphi_{\bar l}
\eea
\end{subequations}
to obtain the following expression, which is one of the main results of this section,
\bea
\label{4dar}
\left( \tilde \bF_{k\bar l} \right)_{I\bar J} &=& \frac{1}{(2\pi)^4} \bigg[
\langle \bar J | \oint \varphi_k
\frac{\left( \d_\tau+2 \right)^4}{\left( \d_\tau+2-x \right)^2} \oint\bar \varphi_l  |I\rangle
- \langle \bar J | \oint \bar \varphi_{\bar l}
\frac{\left( \d_\tau-2 \right)^4}{\left(\d_\tau-2-x \right)^2} \oint\varphi_k  |I\rangle
\nonumber\\
&&-4 \langle \bar J | \oint \varphi_k
\frac{\left( \d_\tau+2 \right)^2}{\left( \d_\tau+2-x \right)^2} \oint\bar \varphi_l  |I\rangle
+4 \langle \bar J | \oint \bar \varphi_{\bar l}
\frac{\left( \d_\tau-2 \right)^2}{\left( \d_\tau-2-x \right)^2} \oint \varphi_k  |I\rangle
\bigg].
\eea

In the limit $x\to 0$ there are obvious cancellations between the denominators and the numerators in this expression. On the r.h.s.\ of the second line some caution is needed as we take the limit. Since there is no contribution from intermediate chiral primary states at any $x\neq 0$, such states need to be subtracted by hand at the $x=0$ expression (precisely as in eq.\ \eqref{2dat} in the 2d $\NN=(2,2)$ case). The subtracted contribution of chiral primary states is proportional to the familiar $[C_k,\bar C_{\bar l}]_{I\bar J}$ term (as a direct 4d analogue of eq.\ \eqref{2dau}).
As a result, 
\bea
\label{4dara}
\left( \bF_{k\bar l} \right)_{I\bar J} = \lim_{x\to 0}  \left( \tilde F_{k\bar l} \right)_{I\bar J} 
= - \left[ C_k, \bar C_{\bar l} \right]_{I\bar J}  + \left( \RR_{k\bar l} \right)_{I\bar J}~,
\eea
where the remainder $\left( \RR_{k\bar l} \right)_{I\bar J}$ is the contact term
\bea
\label{4darb}
\left( \RR_{k\bar l} \right)_{I\bar J} = \frac{1}{(2\pi)^4} \bigg\{
\langle \bar J | \oint \varphi_k 
\left[\left( \d_{\tau}+2 \right)^2-4 \right] \oint\bar \varphi_l  |I\rangle
- \langle \bar J | \oint \bar \varphi_{\bar l}
\left[\left( \d_{\tau}-2 \right)^2-4 \right] \oint\varphi_k  |I\rangle\bigg\}
~.\cr
\eea

\subsection*{Computation of the contact term}

All insertions on the r.h.s.\ of \eqref{4darb} are evaluated at the same time $\tau_1=\tau_2=0$ and when operators come together potential singularities can arise. In order to regularize the expression on the r.h.s.\ we separate the integrated operators in time, setting $\tau_1=-\varepsilon<0$ and $\tau_2=0$, and write
\bea
\label{4darc}
\left( \RR_{k\bar l} \right)_{I\bar J} &=& \frac{1}{(2\pi)^4} \bigg\{
\langle \bar J | \oint \varphi_k (\tau_1)
\left[\left( \d_{\tau_2}+2 \right)^2-4 \right] \oint\bar \varphi_l(\tau_2)  |I\rangle
\cr
&&- \langle \bar J | \oint \bar \varphi_{\bar l}(\tau_1)
\left[\left( \d_{\tau_2}-2 \right)^2-4 \right] \oint\varphi_k(\tau_2)  |I\rangle\bigg\}~.
\eea
At the end of the computation we take the limit $\varepsilon \to 0$. Here we have denoted explicitly the Euclidean time dependence of the (anti)chiral primary field insertions leaving their $S^3$ dependence implicit. 

Since the correlators $\langle \bar J| \oint \phi_k(\tau_1) \oint \bar\varphi_{\bar l}(\tau_2) |I\rangle$ depend only on the difference $\tau_1-\tau_2$ we can turn some of the derivatives $\d_{\tau_2}$ to $-\d_{\tau_1}$. Then, after a few simple algebraic manipulations eq.\ \eqref{4darc} becomes
\bea
\label{4dard}
\left( \RR_{k\bar l} \right)_{I\bar J} &=& -\frac{1}{(2\pi)^4} \bigg\{
e^{4\tau_1}\d_{\tau_1}\d_{\tau_2} 
\left[ e^{-4\tau_1} \langle \bar J | \oint \varphi_k (\tau_1) \oint \bar\varphi_{\bar l}(\tau_2) |I\rangle \right]
\nonumber\\
&&- e^{4\tau_2}\d_{\tau_2}\d_{\tau_1} 
\left[ e^{-4\tau_2} \langle \bar J | \oint \bar \varphi_{\bar l} (\tau_1) \oint \varphi_{k}(\tau_2) |I\rangle \right] 
\bigg \}~.
\eea

Before proceeding with the direct computation of this expression, it is instructive to make the following observation. Equation \eqref{4dard} can be transformed back to the plane using
\beq
\label{4dare}
|x| = e^{\tau_1}~, ~~ |y|=e^{\tau_2}~.
\eeq
Under this transformation the limit $\varepsilon\to 0^+$ translates to the limit $|x| \to 1^-$ with $|y|=1$. Since scalar (anti)chiral primaries $\varphi$ with scaling dimension $\Delta=2$ transform as
\beq
\label{4darf}
\varphi(\tau_1,\vec x) = |x|^2 \varphi(x)~,
\eeq
we find
\bea
\label{4darg}
\left( \RR_{k\bar l} \right)_{I\bar J} &=&- \frac{1}{(2\pi)^4} \bigg\{
|x|^4 \oint \oint (x\cdot \d_x ) (y\cdot \d_y) \left[ \frac{|y|^2}{|x|^2} \langle \bar J | \varphi_k(x) \bar\varphi_{\bar l}(y) | I\rangle \right]
\nonumber\\
&&-|y|^4 \oint \oint (x\cdot \d_x ) (y\cdot \d_y) \left[ \frac{|x|^2}{|y|^2} \langle \bar J |   \bar\varphi_{\bar l}(x) \varphi_k(y) | I\rangle \right] \bigg\}_{|x|\to 1^-, |y|=1}~.
\eea

This form of the contact term is very similar to the form obtained in conformal perturbation theory in \cite{Papadodimas:2009eu} after the use of suitable superconformal Ward identities on the integrated 4-point function formula \eqref{specad} (see eq.\ (C.1) in \cite{Papadodimas:2009eu})
\bea
\label{4darh}
\left( \RR_{k\bar l} \right)_{I\bar J} &=&- \frac{1}{(2\pi)^4} 
\lim_{|y|=1,|x|\to 1^-} \oint \oint |x|^2 |y|^2 (x\cdot \d_x) (y\cdot \d_y)
\nonumber\\
&&\bigg[ 
\frac{|y|^2}{|x|^2} \langle \bar J | \varphi_k(x) \bar\varphi_{\bar l}(y) | I\rangle
- \frac{|x|^2}{|y|^2} \langle \bar J |  \bar\varphi_{\bar l}(x) \varphi_k(y) | I\rangle 
\bigg]~.
\eea

The comparison of the expressions \eqref{4darg} and \eqref{4darh} is very illuminating. The only difference lies on the powers of $|x|$ and $|y|$ outside the integrals; $|x|^4$ and $|y|^4$ in the quantum mechanics formula \eqref{4darg} and the symmetric $|x|^2|y|^2$ in the CFT formula \eqref{4darh}. Since the original expression from quantum mechanics \eqref{4darb} is evaluated at equal zero times $\tau_1=\tau_2=0$, i.e.\ $|x|=|y|=1$, there is no a priori explicit choice for these powers when we write the regularized expression \eqref{4dard} or \eqref{4darg}. Further explicit evaluation of \eqref{4darh}, however, shows that the choice of the external powers is important as we take the limit $\varepsilon \to 0^+$. The choice \eqref{4darg} leads to unreasonable divergences and a non-vanishing Berry phase for the vacuum state. The choice \eqref{4darh} on the other hand, leads to finite results and a vanishing Berry phase for the vacuum state. 
The lesson from this little comparison is that when we write expressions like \eqref{4darg} in quantum mechanics and regulate them, we should be careful about external cutoff-dependent factors. In general such factors should be chosen so that the resulting expression satisfies specific physically motivated properties, e.g.\ that the Berry phase of a single vacuum in a CPT-invariant theory vanishes.

With this lesson in mind we can go back to eq.\ \eqref{4dard} and recast it with the following slight modification of external factors as
\bea
\label{4dari}
\left( \RR_{k\bar l} \right)_{I\bar J} &=&- \frac{1}{(2\pi)^4} \bigg\{
e^{2(\tau_1+\tau_2)}\d_{\tau_1}\d_{\tau_2} 
\left[ e^{-4\tau_1} \langle \bar J | \oint \varphi_k (\tau_1) \oint \bar\varphi_{\bar l}(\tau_2) |I\rangle \right]
\nonumber\\
&&- e^{2(\tau_1+\tau_2)}\d_{\tau_2}\d_{\tau_1} 
\left[ e^{-4\tau_2} \langle \bar J | \oint \bar \varphi_{\bar l} (\tau_1) \oint \varphi_{k}(\tau_2) |I\rangle \right] 
\bigg \}_{\tau_1=-\varepsilon \to 0^-, \tau_2=0}~.
\eea
In analogy to the computation in two dimensions in appendix \ref{2dtechnical} we can also proceed here with a direct computation of this expression on $\IR \times S^3$. Since there are contributions only from regions where the chiral and anti-chiral insertions collide we can evaluate using the OPE between chiral and anti-chiral operators. On the plane, $\IR^4$, the OPE between a chiral and an antichiral operator with scaling dimension $\Delta=2$ takes the form
\beq
\label{4darj}
\varphi({\bf x}_1)\bar \varphi({\bf x}_2) = \sum_{\OO_{\Delta,\ell}} C_{\varphi\bar\varphi}^{\OO_{\Delta,\ell}}
\frac{1}{|{\bf x}_1 - {\bf x}_2|^{4-\Delta}} 
\frac{({\bf x}_1-{\bf x}_2)^{\mu_1}\cdots ({\bf x}_1-{\bf x}_2)^{\mu_\ell}}{|{\bf x}_1- {\bf x}_2|^\ell}
\left(\OO_{\Delta,\ell}\right)_{\mu_1\ldots \mu_\ell}({\bf x}_2)~.
\eeq
In $\IR \times S^3$ coordinates $x^\mu=(\tau, \psi,\theta,\phi)$\footnote{$(\psi,\theta,\phi)$ are the standard $S^3$ spherical coordinates in terms of which the unit $S^3$ metric element takes the form $d\Omega_3^2 = d\psi^2 + \sin^2\psi \left( d\theta^2 + \sin^2\theta d\phi^2\right)$.} this OPE takes the form
\bea
\label{4dark}
&&\varphi(\tau_1, \psi,\theta,\phi)~ \bar \varphi(\tau_2,0,0,0) 
\\
&&=\sum_{\OO_{\Delta,\ell}} C_{\varphi\bar\varphi}^{\OO_{\Delta,\ell}}
\frac{2^{\frac{\Delta}{2}-2} e^{\frac{\Delta}{2}(\tau_1-\tau_2)}}
{\left( \cosh(\tau_1-\tau_2) - \cos\psi \right)^{2-\frac{\Delta}{2}+\frac{\ell}{2}}}
(x_1-x_2)^{\mu_1}\cdots (x_1-x_2)^{\mu_\ell} \left( \OO_{\Delta,\ell}\right)_{\mu_1\cdots \mu_\ell} (\tau_2,0,0,0)~.\nonumber
\eea
Inserting this OPE in \eqref{4dari} one recovers exactly all the steps of the CFT computation in \cite{Papadodimas:2009eu}. The only surviving contributions originate from the conformal blocks of the identity operator, the $\Delta=2, \ell=0$ operator in the supermultiplet that contains the stress-energy tensor, and the $\Delta=3, \ell=1$ operator of the $U(1)_R$ current. The final result is 
\beq
\label{4darl}
\left( \RR_{k\bar l} \right)_{I\bar J} = g_{k\bar l} g_{I\bar J} \left( 1+ \frac{R}{4c}\right)~,
\eeq
where $R$ is the $U(1)_R$ charge of the external states $|I\rangle$, $\langle \bar J|$, and $c$ the central charge
of the $\NN=2$ SCFT.

\subsection*{$tt^*$ equations}

Assembling both contributions in \eqref{4dara} we recover the 4d $\NN=2$ $tt^*$ equations in Ref.\ \cite{Papadodimas:2009eu}
\beq
\label{4das}
\left( \bF_{k\bar l} \right)_{I\bar J} = 
-\left[ C_k,\bar C_{\bar l} \right]_{I\bar J}  +g_{k\bar l} g_{I\bar J} \left( 1+ \frac{R}{4c} \right)~.
\eeq
The above derivation of the \tts equations appears once again to be considerably simpler compared to its conformal perturbation theory counterpart \cite{Papadodimas:2009eu}. This is very encouraging, because while similar results in theories with less symmetry, such as 4d $\NN=1$ SCFTs, are seemingly out of reach in conformal perturbation theory, in the Berry approach of this paper we can easily derive formulae like \eqref{4dara}, \eqref{4darb} even in cases with minimal supersymmetry \cite{inprogress}. This gives us hope that the geometry of the conformal manifold can be analyzed systematically beyond the cases that are currently understood. 

\section{A general relation: Berry versus conformal perturbation theory}
\label{specVScft}

In the previous sections we emphasized the role of the traditional Berry phase, as originally formulated in quantum mechanics, in the context of higher-dimensional quantum field theories. Moreover, in sections \ref{2d}, \ref{4d2} we exhibited the exact equivalence between the Berry curvature of chiral primary states and the curvature of chiral primary operators on conformal manifolds derived independently in conformal perturbation theory. We discussed explicitly the cases of 2d $\NN=(2,2)$ and 4d $\NN=2$ SCFTs, and re-derived the well-known $tt^*$ equations.

In this section we would like to argue that the above equivalence between the Berry connection in quantum mechanics and the connection of conformal perturbation theory holds in general and applies to generic states and operators in any CFT with a non-trivial conformal manifold.
For this purpose, it is useful to begin with a brief review of some of the geometric structures that appear naturally on the conformal manifold of a general $(d+1)$-dimensional CFT from the point of view of conformal perturbation theory. In what follows, we make no assumptions of extra symmetries beyond the standard symmetries related to conformal invariance, e.g.\ we do not assume supersymmetry. 

Let us begin with the main ingredients of the general setup. We consider a $(d+1)$-dimensional CFT on $\IR^{d,1}$ with a non-trivial manifold $\MM$ of exactly marginal deformations parametrized locally by a set of dimensionless couplings $\lambda^\mu$. The manifold $\MM$ is the conformal manifold of the CFT. On the level of an action\footnote{In what follows we do not assume a specific Lagrangian formulation of the CFT. An action is invoked here for illustrational purposes to express the operators that define the tangent space of the conformal manifold.} $\SS$, an infinitesimal deformation across $\MM$ takes the form
\beq
\label{specaa}
\delta \SS = \delta \lambda^\mu \int d^{d+1}x ~ \OO_\mu(x)~,
\eeq
where $\OO_\mu$ is an operator with exact scaling dimension $\Delta = d+1$. It will be convenient to work in the Wick rotated Euclidean version of the CFT.

The set of local operators of the CFT (at any scaling dimension) defines a formal infinite-dimensional vector bundle of local operators 
\beq
\label{specab}
\BB_{operator} ~\longrightarrow ~ \MM 
\eeq
over the conformal manifold $\MM$. At each point $\{ \lambda^\mu \}$ of the base conformal manifold $\MM$ the fiber is the vector space of local operators $\OO(x)$ in the CFT defined at $\{ \lambda^\mu \}$. 
The generic section of this bundle describes a $\lambda$-dependent basis of local operators of the CFT. As one traces a curve on the conformal manifold, operators of the same scaling dimension mix. This mixing, which is an inherent property of the quantum dynamics of the CFT, is encoded naturally in a non-trivial connection on the vector bundle $\BB_{operator}$. Once the notion of a connection is available the comparison of two operators at nearby points of the conformal manifold becomes feasible and standard geometric notions, like that of a covariant derivative $\nabla_\mu$ and parallel transport, immediately apply.

A natural definition of such a connection in conformal perturbation theory, which follows directly from the dynamics of the CFT, has been discussed in many works in the past (see for example \cite{Kutasov:1988xb} for an early discussion, \cite{Ranganathan:1993vj} for an extensive discussion in two-dimensional CFTs, as well as \cite{Friedan:2012hi}). The curvature of this connection (denoted $\AA_\mu$)
\beq
\label{specac}
\left(F_{\mu\nu}\right)_{IJ} = \left[ \nabla_\mu, \nabla_\nu \right]_{IJ}~,~~  \nabla_\mu = \d_\mu +\AA_\mu
\eeq
can be expressed in CFT in terms of the integrated 4-point function
\beq
\label{specad}
\left(F_{\mu\nu}\right)_{IJ} = \int_{|x|\leq 1} d^{d+1}x \int_{|y|\leq 1} d^{d+1}y  ~
\langle \OO_J(\infty) \OO_{[\mu}(x) \OO_{\nu]}(y) \OO_I(0) \rangle~.
\eeq
In this formula the exactly marginal operators $\OO_\mu$, $\OO_\nu$ are integrated in a unit ball on the $(d+1)$-dimensional plane, and the arbitrary external operators $\OO_I$, $\OO_J$ are inserted at the origin and infinity respectively.

As usual in QFT, the collision of two operators in a correlation function leads to potential ultraviolet (UV) divergences that need to be regularized. In general, different regularization schemes lead to different notions of connection on the conformal manifold. In \cite{Ranganathan:1993vj} several possibilities were discussed in detail. One of them requires the introduction of small cutoff balls around the operator insertions so that two operators can never collide in the regulated expression. At the end of the computation, the cutoff size is sent to zero and divergent terms are removed by hand. This is the prescription that is implicitly used in \eqref{specad}.

At this point, it should be clear that the objects defined in eqs.\ \eqref{specac}, \eqref{specad} are conceptually close to the notions of Berry phase and Berry connection in quantum mechanics, as they were outlined in the previous sections. In both cases, one discusses how physical quantities vary under the adiabatic changes of parameters in the theory. 
By invoking the operator-state correspondence in CFT it is possible to make this relation much more explicit, generalizing the results of the previous two subsections.

The operator-state correspondence arises naturally in radial quantization. Equivalently, with a standard conformal transformation the (Wick-rotated) flat space theory transforms to the theory on the hyper-cylinder $\IR_\time \times S^d$. In this context, we have a natural formulation of the CFT as a one-dimensional quantum mechanics theory in terms of a Hamiltonian $H$ whose spectrum measures the scaling dimension of different states. The original dependence of the CFT on the couplings $\lambda^\mu$ translates to a $\lambda$-dependent Hamiltonian. Hence, by considering adiabatic changes of the couplings $\{ \lambda^\mu \}$ one is led to the Berry phase of states and the corresponding Berry-Simon connection on the vector bundle of Hilbert spaces of states
\beq
\label{specae}
\BB_{states} ~\longrightarrow ~ \MM
\eeq
over the conformal manifold $\MM$. The correspondence between states and operators, implemented by local operators acting at the origin of the plane (or equivalently by operators acting at $\time=-\infty$ on the cylinder)\footnote{As we did in sections \ref{2d}, \ref{4d2}, it is in fact convenient to use a closely related basis of states obtained from \eqref{specaf} by a similarity transformation. The details of this transformation are summarized in appendix \ref{conventions}. We denote the state obtained in this way from $|\OO\rangle_I$ as $|I\rangle$.}
\beq
\label{specaf}
|\OO\rangle_I = \OO_I(0) |0\rangle~,
\eeq
guarantees a map between connections and holonomies on $\BB_{states}$ and $\BB_{operators}$. Under this map the Berry connection maps to a corresponding connection in conformal perturbation theory. We will claim that this connection is naturally the one leading to the curvature \eqref{specad}.

With these specifications, Berry's prescription provides a connection with components
\beq
\label{specaga}
\left( \bA_\mu \right)_{IJ} = \langle J | \d_\mu | I\rangle~.
\eeq 
As we recalled in section \ref{BerryReview}, and appendix \ref{spectralBerry}, the curvature of this connection can be expressed quite generally as a spectral sum of the form
\beq
\label{specag}
\left( \bF_{\mu\nu} \right)_{IJ}
=  \sum_{n \not\in \HH_I} \sum_{a,b,\in \HH_n} 
\frac{1}{(\Delta_I-\Delta_n)^2} \langle J | \p_\mu H | n,a \rangle g_{(n)}^{ab} \langle n,b | \p_\nu H | I \rangle 
- (\mu \leftrightarrow \nu)~,
\eeq
where $\Delta_n$ is the scaling dimension, i.e.\ energy, of a state $|n\rangle$ in the Hilbert subspace $\HH_n$.

We have seen in previous sections in explicit evaluations of the r.h.s.\ of equation \eqref{specag} applied to CFTs, that this formula
is typically divergent and, like \eqref{specad}, it requires a regularization prescription.  

We can now ask the central question of this section: does the operator-state correspondence imply a precise relation between the quantity defined in \eqref{specag} and the CFT 4-point function formula \eqref{specad}? To answer this question, it is first convenient to observe that \emph{the Berry curvature is independent of terms in $\d_\mu H$ that commute with the Hamiltonian}.

To see this, let us write the Hamiltonian derivatives $\d_\mu H$ in the form
\beq
\label{specai}
\d_\mu H = \HH_\mu + \RR_\mu~,
\eeq
with $\HH_\mu$ arbitrary but $\RR_\mu$ having the property
\beq
\label{specaj}
[H, \RR_\mu]=0~.
\eeq
Then, for $\Delta_n \neq 0$ (namely, $|n\rangle$ different from the ground state $|0\rangle$) 
\beq
\label{specak}
\langle J | \RR_\mu |n\rangle 
= \frac{1}{\Delta_n} \langle J | \RR_\mu H |n\rangle
= \frac{1}{\Delta_n} \langle J | H \RR_\mu  |n \rangle
= \frac{\Delta_J}{\Delta_n} \langle J |\RR_\mu | n\rangle~.
\eeq
Assuming $\Delta_J \neq \Delta_n$, as is the case with all terms in \eqref{specag}, we deduce $\langle J |\RR_\mu |n \rangle=0$ and therefore
\beq
\label{specal}
\left( \bF_{\mu\nu} \right)_{IJ}
= \sum_{n\not\in \HH_I} \sum_{a,b\in \HH_n} 
\frac{1}{(\Delta_I-\Delta_n)^2} \langle J | \HH_\mu | n,a \rangle g_{(n)}^{ab} \langle n,b | \HH_\nu | I \rangle 
- (\mu \leftrightarrow \nu)
\eeq
is independent of $\RR_\mu$.

If the external states are the vacuum $|0\rangle$, the states $|n\rangle$ over which we sum in \eqref{specag} cannot be ground states, hence \eqref{specak} applies as it is. If the external states are not the vacuum, and the vacuum contributes to the sum \eqref{specag}, then we can still deduce $\langle J |\RR_\mu |0 \rangle=0$ by writing $\langle J |\RR_\mu |n \rangle= \frac{1}{\Delta_J} \langle J |H \RR_\mu |n \rangle= \langle J |\RR_\mu H |n \rangle =0$, which leads to the desired result.

In our case, the Hamiltonian deformations $\d_\mu H$ are operators at $\time=0$ integrated over the sphere $S^d$
\beq
\label{specam}
\d_\mu H = \int_{S^d}d^d z \sqrt{g_{S^d}} ~ \OO_\mu(0,z) 
\equiv \oint \OO_\mu(0)~,
\eeq
where $\OO_\mu(t,z)$ represents the exactly marginal interaction $\OO_\mu$ on $\IR\times S^d$. 
To keep the notation brief we indicate the integral over the $d$-dimensional round sphere as $\oint$ and keep only the time dependence explicit (in \eqref{specam} $\tau=0$).
Then, by a simple integration by parts, and using the fact that $[H, \OO]=\d_\time \OO$, we observe that we can write
\beq
\label{specan}
\oint \OO_\mu(0) 
= [H, \CC_\mu] + \DD_\mu~,
\eeq
where
\beq
\label{specao}
\CC_\mu =  -\int_{0}^\infty d\tau \oint \OO_\mu(\tau)
~,~~
\DD_\mu = \oint \OO_\mu(\infty)~.
\eeq
Since $\OO_\mu$ represents an exactly marginal deformation, it commutes with the Hamiltonian when inserted at $\tau=\infty$ (or equivalently at the asymptotic infinity in flat space). Hence, exact marginality implies
\beq
\label{specap}
[H,\DD_\mu]=0~.
\eeq 

As a result, by combining \eqref{specan}-\eqref{specap} with the above lemma we learn that we can recast the Berry curvature \eqref{specag} into the form
\beq
\label{specaq}
\left( \bF_{\mu\nu} \right)_{IJ}
= \sum_{n\not\in \HH_I} \sum_{a,b\in \HH_n}
\frac{1}{(\Delta_I-\Delta_n)^2} \langle J | [H,\CC_\mu] | n,a \rangle g_{(n)}^{ab} \langle n,b | [H,\CC_n] | I \rangle 
- (\mu \leftrightarrow \nu)~,
\eeq
which implies trivially
\beq
\label{specar}
\left( \bF_{\mu\nu} \right)_{IJ}
= - \sum_{n\not\in \HH_I} \sum_{a,b\in \HH_n}
\langle J | \CC_\mu | n,a \rangle g_{(n)}^{ab}  \langle n,b | \CC_\nu | I \rangle 
- (\mu \leftrightarrow \nu)~.
\eeq
Adding and subtracting the sum over states with scaling dimension $\Delta_I$ in the Hilbert subspace $\HH_I$ of the external states we further obtain
\bea
\label{specas}
\left( \bF_{\mu\nu} \right)_{IJ}
=&& - \int^{\infty}_0 d\time \int^{\infty}_0 d\time' 
\langle J | \left[ \oint \OO_\mu (\time), \oint\OO_\nu(\time') \right] | I\rangle
\nonumber\\
&&+ \sum_{M,N\in \HH_I}
\langle J | \CC_\mu | M \rangle g^{MN} \langle N | \CC_\nu | I \rangle 
- (\mu \leftrightarrow \nu)~.
\eea

Interestingly, the second line on the r.h.s.\ of eq.\ \eqref{specas} does not contribute. Indeed, the second line, which is
\beq
\label{specasa}
\int^{\infty}_0 d\time \int^{\infty}_0 d\time' \sum_{M,N\in \HH_I} \langle J | \d_\mu H | M \rangle g^{MN} 
\langle N | \d_\nu H | I \rangle  - (\mu\leftrightarrow \nu)
\eeq
can be evaluated using the identity 
\beq
\label{specasb}
\langle J | \d_\mu H | M \rangle =  \d_\mu \Delta \, g_{JM}
\eeq
to obtain 
\beq
\label{specasc}
\int^{\infty}_0 d\time \int^{\infty}_0 d\time' \d_\mu \Delta\, \d_\nu \Delta\, g_{IJ} - (\mu \leftrightarrow \nu) 
\eeq
which vanishes by anti-symmetry. The identity \eqref{specasb} can be proved easily by taking the $\mu$-derivative of $\langle J | H | M \rangle = \Delta g_{JM}$ ($\Delta$ being the scaling dimension in the Hilbert subspace $\HH_J$). Notice that each of the factors $\langle J |\d_{\mu} H | M\rangle$ vanishes identically if the deformation does not change the scaling dimension $\Delta$.
This is what happened with the chiral primary external states in the examples analyzed in the previous sections \ref{2d}, \ref{4d2}.

Consequently, the final form of eq.\ \eqref{specas} is 
\beq
\label{specasd}
\left( \bF_{\mu\nu} \right)_{IJ}
= - \int^{\infty}_0 d\time \int^{\infty}_0 d\time' 
\langle J | \left[ \oint \OO_\mu (\time), \oint\OO_\nu(\time') \right] | I\rangle~.
\eeq
A time reversal transformation, $\tau \to -\tau$, together with a conformal transformation of this equation back to the plane yields the 4-point function expression in \eqref{specad}. This establishes the general formal equivalence of the expressions \eqref{specag}, \eqref{specad}. 

\section{Discussion}
\label{open}

In this paper we discussed general aspects of the Berry phase in QFT. We showed that a non-trivial Berry phase emerges already in very simple quantum field theories, such as free electromagnetism with a theta angle. In this case, as we adiabatically vary the EM couplings $\g$ and $\theta$, the polarization vector of a linearly polarized photon rotates in the plane orthogonal to its momentum. Therefore, this effect is potentially measurable in materials where the effective electromagnetic couplings can be manipulated. We hope to analyze this possibility in greater detail in a future publication.

It would be interesting to extend the results presented in this paper to further computable cases and to study their physical implications. An obvious possibility is to study the Berry phase of BPS states in more general supersymmetric theories. For example, in the context of 4d ${\cal N}=1$ theories, it is natural to conjecture, extending the results of section \ref{sec:n2coulomb}, that the Riemann tensor on the moduli space of vacua characterizes the Berry phase of massless scalars as we move on the moduli space.
Another especially interesting case is the Berry phase of chiral primary states in 4d ${\cal N}=1$ SCFTs, which we plan to address in future work \cite{inprogress}. Extensions to massive ${\cal N}=2$ theories are also worth investigating further.

Motivated by the observation that the Berry phase of low-energy states in the Coulomb branch of $\NN=2$ theories is determined by the Riemann tensor, it appears natural to conjecture that a similar result should hold for supersymmetric compactifications in string theory. The Riemann tensor on these moduli spaces can be related to a certain combination of low energy $2\rightarrow 2$ scattering amplitudes \cite{Dixon:1989fj}, where two of the states are the particles whose Berry phase we want to compute and the other two are the moduli along which we are computing the Berry curvature tensor. It might be interesting to explore whether the Berry phase of massive string states and D-branes could be related to the low energy limit of an S-matrix of moduli scattered off the massive states.

In some of the computations in this paper, we introduced a compact spatial manifold to deal with infrared divergences, and showed that the results survive in the decompactification limit. It would be extremely interesting to study the Berry phase for quantum field theories defined on more general compact manifolds, where it could potentially provide new interesting observables.

Finally, in this paper we considered the Berry phase only in local patches of the parameter space. It would be interesting to investigate global aspects over the parameter space (see \cite{Kiritsis:1986re} for a discussion of global properties of the Berry phase).

\subsection*{Acknowledgments}

We would like to thank Costas Bachas, Alex Bols, Atish Dabholkar, Aristos Donos, Daniele Dorigoni, Jaume Gomis, Elias Kiritsis, Zohar Komargodski, Edoardo Lauria, David Tong, and especially Cumrun Vafa. The work of MB was supported in part by the European Research Council grant no.~ERC-2013-CoG 616732 HoloQosmos and in part by the FWO and the European Union's Horizon 2020 research and innovation programme under the Marie Sk{\l}odowska-Curie grant agreement no.~665501. MB is an FWO [PEGASUS]${}^2$ Marie Sk{\l}odowska-Curie Fellow. KP would like to thank the ENS, Paris for hospitality while this paper was being completed and the Royal Netherlands Academy of Sciences (KNAW). Preliminary results based on this work were presented by VN in the workshop String Theory in London, Aug.\ 29-Sept.\ 3 2016 at King's College, London.

\begin{appendix}

\section{Spectral formula for non-abelian Berry curvature in quantum mechanics}
\label{spectralBerry}

In this appendix we summarize, for the benefit of the reader, a quick derivation of the spectral QM formula for the non-abelian Berry curvature \eqref{specag}. This is one of the main formulae used in the main text. 

Recall that the general non-abelian Berry (or Wilczek-Zee) connection has components
\beq
\label{nonabelaa}
\left( \bA_\mu^{(n)} \right)_{ab}  =  \langle n,b | \p_\mu | n, a \rangle~,
\eeq
where we use labels $a,b,\ldots = 1,\ldots, N_n$ to label the degeneracy for the states in the degenerate sector $\HH_n$. The corresponding curvature is 
\beq
\label{nonabelab}
\left( \bF_{\mu\nu}^{(n)} \right)_{a}^{~b} 
= \p_\mu \left( \bA_\nu^{(n)} \right)_{a}^{~b} - \p_\nu \left( \bA_\mu^{(n)} \right)_{a}^{~b}
- \left[ \bA_\mu^{(n)}, \bA_\nu^{(n)} \right]_{a}^{~b}~.
\eeq
Lowering the upper index $b$ with the metric (matrix of 2-point functions) $g_{(n)ab}=\langle n,a | n,b\rangle$ in the eigenspace with eigenvalue $E_n$ we get
\bea
\label{nonabelaba}
\left( \bF_{\mu\nu}^{(n)} \right)_{ab} 
&=& \p_\mu \left( \bA_\nu^{(n)} \right)_{ab} - \p_\nu \left( \bA_\mu^{(n)} \right)_{ab}
-  \left[ \bA_\mu^{(n)}, \bA_\nu^{(n)} \right]_{ab} 
\cr
&&+ \left( \bA_\nu^{(n)} \right)_{ad} \d_\mu g_{(n)}^{dc} g_{(n)cb}
- \left( \bA_\mu^{(n)} \right)_{ad} \d_\nu g_{(n)}^{dc} g_{(n)cb}~.
\eea
Hence, in a more explicit form for the first three terms of the r.h.s.\ of this equation 
\bea
\label{nonabelac}
&&\left( \bF_{\mu\nu}^{(n)} \right)_{ab}
= \langle \p_\mu (n,b) |  \p_\nu (n,a) \rangle - \langle \p_\nu (n,b) |  \p_\mu (n,a) \rangle
\nonumber\\
&&- \sum_{c,d} \langle (n,c) |  \p_\mu (n,a) \rangle g_{(n)}^{cd} \langle (n,b) |  \p_\nu (n,d) \rangle
+\sum_{c,d} \langle (n,c) |  \p_\nu (n,a) \rangle g_{(n)}^{cd} \langle (n,b) |  \p_\mu (n,d) \rangle 
\nonumber\\
&&+ \left( \bA_\nu^{(n)} \right)_{ad} \d_\mu g_{(n)}^{dc} g_{(n)cb}
- \left( \bA_\mu^{(n)} \right)_{ad} \d_\nu g_{(n)}^{dc} g_{(n)cb}~.
\eea
Inserting the identity
\beq
\label{nonabelad}
1 = \sum_{m,c,d} |m,c\rangle g_{(m)}^{cd}  \langle m,d |
\eeq
in the first line of \eqref{nonabelac} we obtain 
\bea
\label{nonabelae}
&&\left( \bF_{\mu\nu}^{(n)} \right)_{ab}
= \sum_{m,c,d} \left( 
\langle \p_\mu (n,b) | m,c \rangle g_{(m)}^{cd} \langle m,d| \p_\nu (n,a) \rangle - 
\langle \p_\nu (n,b) | m,c\rangle g_{(m)}^{cd} \langle m,d| \p_\mu (n,a) \rangle \right)
\nonumber\\
&&- \sum_{c,d} \left( \langle (n,c) |  \p_\mu (n,a) \rangle g_{(n)}^{cd} \langle (n,b) |  \p_\nu (n,d) \rangle
-\langle (n,c) |  \p_\nu (n,a) \rangle g_{(n)}^{cd} \langle (n,b) |  \p_\mu (n,d) \rangle \right)
\nonumber\\
&&+ \left( \bA_\nu^{(n)} \right)_{ad} \d_\mu g_{(n)}^{dc} g_{(n)cb}
- \left( \bA_\mu^{(n)} \right)_{ad} \d_\nu g_{(n)}^{dc} g_{(n)cb}~.
\eea
A simple computation shows that for $m\neq n$
\beq
\label{nonabelaf}
\langle m,c | \p_\mu (n,a) \rangle = \frac{ \langle m,c | \p_\mu H | n,a \rangle }{E_n-E_m}~.
\eeq
Then, inserting \eqref{nonabelaf} into \eqref{nonabelae} we obtain
\bea
\label{nonabelag}
\left( \bF_{\mu\nu}^{(n)} \right)_{ab}
&=&  \Big[ \sum_{m\neq n, c,d} \frac{1}{(E_n-E_m)^2}  \langle n,b | \p_\mu H | m,c \rangle
g_{(m)}^{cd}
\langle m,d | \p_\nu H | n,a \rangle
\cr
&& +\sum_{c,d} \langle \p_\mu(n,b) | n,c \rangle g_{(n)}^{cd} \langle n,d | \p_\nu (n,a) \rangle
\cr
&& +\sum_{c,d} \langle n,d | \p_\nu(n,a) \rangle  g_{(n)}^{cd} 
\langle n,b | \p_\mu (n,c) \rangle \Big] - (\mu \leftrightarrow \nu)
\cr
&&+ \left( \bA_\nu^{(n)} \right)_{ad} \d_\mu g_{(n)}^{dc} g_{(n)cb}
- \left( \bA_\mu^{(n)} \right)_{ad} \d_\nu g_{(n)}^{dc} g_{(n)cb}
\cr
&=&  \sum_{m\neq n, c,d} \frac{1}{(E_n-E_m)^2}  \langle n,b | \p_\mu H | m,c \rangle 
g_{(m)}^{cd}
\langle m,d | \p_\nu H | n,a \rangle  
- (\mu \leftrightarrow \nu)
\cr
&& +\left( \bA_\nu^{(n)} \right)_{ad} g_{(n)}^{cd} \d_\mu g_{(n)bc} 
- \left( \bA_\mu^{(n)} \right)_{ad} g_{(n)}^{cd} \d_\nu g_{(n)bc} 
\cr
&&+ \left( \bA_\nu^{(n)} \right)_{ad} \d_\mu g_{(n)}^{dc} g_{(n)cb}
- \left( \bA_\mu^{(n)} \right)_{ad} \d_\nu g_{(n)}^{dc} g_{(n)cb}~.
\eea
The last two lines in \eqref{nonabelag} obviously cancel out.

As a result, we obtain the final formula
\beq
\label{nonabelak}
\left( \bF_{\mu\nu}^{(n)} \right)_{ab}
=  \sum_{m\neq n,c,d} 
\frac{1}{(E_n-E_m)^2} \langle n,b | \p_\mu H | m,c \rangle g_{(m)}^{cd} \langle m,c | \p_\nu H | n,a \rangle 
- (\mu \leftrightarrow \nu)
~.
\eeq
Typically this result is quoted in a set of orthonormal intermediate states where $g_{(m)}^{cd}=\delta^{cd}$ with common emphasis on the abelian case.

\section{Berry phase in systems with anti-unitary symmetries}
\label{antiproof}

Suppose that the system is invariant under an anti-unitary symmetry $\au$. By this we mean that there is a {\it fixed} anti-linear operator obeying $\au^\dagger \au=1$ and
\beq
\label{antiaa}
[H(\lambda),\au] = 0\qquad \forall \lambda~.
\eeq
In the case where $\au$ is time reversal or CPT, it additionally obeys $\au^2 = \pm 1$. We will analyze the consequences of this symmetry on the Berry phase by starting with the abelian case and then proceeding with the non-abelian one.

\vskip10pt
\noindent {\bf Abelian case.} Here we consider an energy eigenstate $|n\rangle$ that is nondegenerate. The non-degeneracy implies that on this state we must have $\au^2=1$.\footnote{Given that there are no degeneracies we must have $\au |n\rangle = e^{i \phi} |n\rangle$. Then, we have $\au^2 |n\rangle = \au (e^{i\phi} |n\rangle) = e^{-i\phi} \au |n\rangle = |n\rangle$. So $\au^2=1$ when acting on a non-degenerate state.}
With a suitable choice of the phase of the state we can ensure that over an open neighborhood of the parameter space we have
\beq
\label{antiab}
\Theta |n(\lambda)\rangle = |n (\lambda)\rangle~.
\eeq
Notice that this immediately implies
\beq
\label{antiac}
\Theta \partial_i |n(\lambda)\rangle = \partial_i |n (\lambda)\rangle~.
\eeq

With these specifications we observe that
\beq
\label{antiad}
A_i = \langle n| \partial_i |n\rangle=\langle n| \au^\dagger \au \partial_i |n\rangle = \langle n| \au^\dagger  \partial_i |n\rangle=\langle \partial_i  n| \au |n\rangle
= \langle \partial_i n | n\rangle = A_i^*~.
\eeq
At the same time, from the fact that $A_i$ corresponds to an anti-Hermitian connection we have the very basic property
\beq
\label{antiae}
0=\partial_i (\langle n |n\rangle ) = \langle \partial_i n |n\rangle + \langle n|\partial_i |n\rangle = A_i^* + A_i~.
\eeq
Combining the last two equations we find $A_i=0$. 

This shows that if an energy eigenstate is non-degenerate in a system with $\au$-invariance, then the Berry phase for this state must be equal to zero. A general implication of this result, which was emphasized in the main text, is the following. 
Relativistic QFTs are CPT-invariant. If there is also a unique ground state, then the Berry phase associated to it should be zero. This results holds even for a QFT defined on a manifold of the form $\IR \times {\cal T}$, provided that CPT-invariance remains true and that the ground state is unique.

\vskip10pt
\noindent {\bf Non-abelian case.} More generally, suppose we have a subspace of degenerate states $|n,a\rangle$ $a=1,...N$, where the operator $\au$ acts accordingly. We will consider two possibilities: $i)$ $\au^2 = 1$, or $ii)$ $\au^2 = -1$ on this subspace. 

\vskip10pt
$i)$ \underline{$\au^2=1$}: A simple linear algebra argument shows that we can select an orthonormal basis of states on this subspace obeying
\beq
\label{antiba}
\au |n,a\rangle(\lambda) = |n,a \rangle(\lambda)~~, \qquad \langle n,a | n,b\rangle = \delta_{ab}~.
\eeq
Writing \eqref{revba} in this basis we find
\bea
\label{antibb}
(\bA^{(n)}_{i})_{ab} &=& \langle n,b| \partial_i |n,a \rangle  
=  \langle n,b| {\au^\dagger \au}\partial_i |n,a \rangle  
=  \langle n,b| {\au^\dagger}\partial_i |n,a \rangle
\cr
&=& (\partial_i \langle n,a|) \au |n,b\rangle = (\partial_i \langle n,a |) |n,b\rangle = \left(\langle n,b| \partial_i |n,a\rangle \right)^* = (\bA^{(n)}_{i})_{ab}^*~,
\eea
which means that the connection matrix $\bA^{(n)}$ is not only anti-Hermitian, but, moreover, that there is a basis in a local neighborhood where the matrix elements are real. This implies that the vector bundle has reduced holonomy from $U(N)$ down to $O(N)$.

\vskip10pt
$ii)$ \underline{$\au^2=-1$}: A first observation in this case is that the subspace must have an even dimension $N=2k$. Again, a simple linear algebra argument shows that we can select a basis of states consisting of $k$  states $|i\rangle$, as well as their images under $\au$ defined as  $|\widetilde{i}\rangle\equiv\au |i\rangle$, $i=1,,,.k$. The $N=2k$ states $|i\rangle, |\widetilde{i}\rangle$ provide an orthonormal basis, and they have simple transformation under $\au$, namely
\beq
\label{antica}
\au |i\rangle = |\widetilde{i}\rangle~~, \qquad \au |\widetilde{i}\rangle = - |i\rangle~.
\eeq
In this basis the Berry connection takes the form
\beq
(\bA^{(n)}_{\mu})_{ij} = \langle j| \partial_\mu |i \rangle  =  (\au \partial_\mu |i\rangle, \au |j\rangle) 
= ( \partial_\mu |\widetilde{i}\rangle, |\widetilde{j}\rangle) 
= - (|\widetilde{i}\rangle,\partial_\mu |\widetilde{j}\rangle) = - (\bA^{(n)}_{\mu})_{\widetilde{j} \,\widetilde{i}}~,
\eeq
\beq
(\bA^{(n)}_{\mu})_{\widetilde j i} = \langle i | \partial_\mu | \tilde j \rangle  
=  \langle i | \au \partial_\mu |j\rangle 
= (\au \au \partial_\mu |j\rangle, \au |i\rangle)
= -(\partial_\mu |j\rangle, | \tilde{i}\rangle) 
= \langle j| \partial_\mu |\widetilde{i}\rangle = (\bA^{(n)}_{\mu})_{\widetilde i j}~,
\eeq
and similarly we can show that $(\bA^{(n)}_{\mu})_{j \widetilde{i}} = (\bA^{(n)}_{\mu})_{i \widetilde{j} }$. So if we think of the connection matrix as a $(2k) \times (2k)$ matrix consisting of 4 $k\times k$ blocks,  we find that the matrix has the form 
\beq
\left( \begin{matrix}
  A & B \\
  C & -A^T \end{matrix}\right)~,
\eeq
where $B,C$ are symmetric. This is the condition for an $Sp(N)$ connection.

\section{Derivation of photon Berry phase in electromagnetism}
\label{detailsqed}

In this appendix we consider the theory of electromagnetism with a theta-term interaction. The Lagrangian is
\beq
\label{Maxa}
\LL = - \frac{1}{4\g^2} F_{\mu\nu} F^{\mu\nu} + \frac{\theta}{32\pi^2} F_{\mu\nu} \tilde F^{\mu\nu}~,
\eeq
or in terms of the electric and magnetic fields ($E^i = - F^{0i}$, $B^i = \varepsilon^{ijk}F_{ij}$ respectively)
\beq
\label{Maxaa}
\LL = \frac{1}{2\g^2} \left( \vec E^2 - \vec B^2 \right) - \frac{\theta}{8\pi^2} \vec E\cdot \vec B~.
\eeq

The canonical momentum $\vec \pi$ conjugate to the vector potential $\vec A$ has components
\beq
\label{Maxad}
\pi_i = \frac{\d \LL}{\d \d_t A_i} = \frac{1}{\g^2} E_i -\frac{\theta}{8 \pi^2} B_i
\eeq
and the momentum $\pi^0= \frac{\d \LL}{\d \d_t A_0}$ vanishes as a first class constraint.  The Hamiltonian takes the form
\beq
\label{Maxac}
H = \frac{1}{2} \int d^3x \left[ \g^2 \left( \vec \pi + \frac{\theta}{8\pi^2} \vec B \right)^2 + \frac{1}{\g^2} \vec B^2
+ \vec \pi \cdot \vec \nabla A_0  \right]~.
\eeq
Consequently, its derivatives with respect to the couplings $\g^2$, $\theta$ are
\beq
\label{Maxae}
\d_{\g^2} H
= \frac{1}{\g^4}\int d^3x\, \left( \vec E^2- \vec B^2\right)~, ~~
\d_\theta H 
= \frac{1}{8\pi^2} \int d^3x \, \vec E \cdot \vec B~.
\eeq

We assume that the three space directions are compactified on a $\IT^3$ with, say, common size $R$ and volume $V=R^3$. When $\theta$ is constant in the absence of physical boundaries the $\theta$-interaction in \eqref{Maxa} is a total derivative that does not affect the equations of motion. Nevertheless, as we see explicitly in \eqref{Maxae} the variation of $H$ with respect to $\theta$ can be non-zero and eventually will contribute non-trivially to the Berry phase computation. Since 
\beq
\label{Maxaea}
\d_\theta H \propto \int d^3 x~ \vec E \cdot \vec B = 
\frac{1}{2} \int d^3x~ \varepsilon_{ijk} \Big[ \d_j \left( \d_t A_i A_k\right) 
-\d_t \left( \d_j A_i A_k \right) \Big]~,
\eeq
it is the second term on the r.h.s., which is a total derivative in time, that is expected to contribute. This fits nicely with the fact that, eventually, we consider effects associated to the adiabatic change of $\theta$ in time. 

With these specifications we proceed to evaluate the Berry curvature of photon states using the general equation \eqref{berrycurv}.

In standard fashion we quantize the theory in the Coulomb gauge, where $A_0=0, \vec \nabla \cdot \vec A=0$. In this gauge the vector gauge potential can be expanded in creation and annihilation modes with two possible helicities
\beq
\label{Maxaf}
\vec A (t, \vec x) =  \sum_{\vec k} \sum_{\epsilon=\pm} \sqrt{\frac{\hhbar \g^2}{2 \omega_k V}}
\Big( \vec e_\epsilon (\vec k) a_{\vec k ,\epsilon} e^{-i\omega_k t + i \vec k \cdot \vec x}
+  \vec {\bar e}_\epsilon (\vec k) a^\dagger_{\vec k ,\epsilon} e^{i\omega_k t -i \vec k \cdot \vec x} \Big )~.
\eeq
In units where $c=1$, $\omega_k=|\vec k|$ denotes the frequency of the modes. The spatial momenta are quantized on $\IT^3$ as $k_i = \frac{2\pi n_i}{R}$, for $n_i \in \IZ$, and $i=1,2,3$. $\epsilon=\pm $ are the two helicities of the photon modes and $\vec e_\epsilon (\vec k)$ the polarization vectors.\footnote{Possible global issues in defining the polarization vectors for arbitrary $\vec{k}$ will not play a role in the following.} The creation and annihilation modes $a$, $a^\dagger$ obey canonical commutation relations.

The corresponding expansion of the electric and magnetic fields is
\beq
\label{Maxag}
\vec E = i  
\sum_{\vec k} \sum_{\epsilon=\pm} \sqrt{\frac{\hhbar \g^2 \omega_k}{2 V}} 
\Big(
\vec e_{\epsilon}(\vec k)  a_{\vec k ,\epsilon} e^{-i\omega_k t + i \vec k \vec x} 
- \vec{\bar e}_{\epsilon} (\vec k) a^\dagger_{\vec k,\epsilon} e^{i \omega_k t-i \vec k \vec x} 
\Big)~,
\eeq
where we used the fact that in Coulomb gauge $\vec{k} \cdot \vec e_\epsilon (\vec k)=0$, and
\beq
\label{Maxai}
\vec B =   i  
\sum_{\vec k} \sum_{\epsilon=\pm} \sqrt{\frac{\hhbar \g^2}{2\omega_k V}} 
\Big(
(\vec k \times \vec e_{\epsilon}(\vec k))  a_{\vec k ,\epsilon} e^{-i\omega_k t + i \vec k \vec x} 
- (\vec k \times \vec{\bar e}_{\epsilon} (\vec k)) a^\dagger_{\vec k,\epsilon} e^{i \omega_k t-i \vec k \vec x} 
\Big)~.
\eeq
Evaluating the Hamiltonian derivatives at $t=0$, we find after some straightforward algebra
\beq
\label{Maxaj}
\d_{\g^2} H = - \frac{1 \hhbar}{2\g^3} \sum_{\vec k}\sum_{\epsilon=\pm} \omega_k \left( 
a_{\vec k,\epsilon} a_{-\vec k,\epsilon} 
+a^\dagger_{\vec k,\epsilon} a^\dagger_{-\vec k,\epsilon} \right)~,
\eeq 
\beq
\label{Maxak}
\d_\theta H = \frac{i \hhbar \g^2}{16\pi^2} \sum_{\vec k}\sum_{\epsilon=\pm} 
\omega_k \left( \epsilon \left( a_{\vec k,\epsilon} a_{-\vec k,\epsilon} - a^\dagger_{\vec k,\epsilon} a^\dagger_{-\vec k,\epsilon} \right) 
+ \left( a_{\vec k,\epsilon} a^\dagger_{\vec k,\epsilon} - a^\dagger_{\vec k,\epsilon} a_{\vec k,\epsilon} \right) \right)~.
\eeq

When we evaluate the Berry curvature $F_{\g^2\theta}^{(n)}$ in a state $|n\rangle$ we encounter terms of the form $\langle n | \d_{\g^2} H | m\rangle \langle m | \d_\theta H | n\rangle$. Clearly, terms in $\delta_\theta H$ of the form $a^\dagger a$ or $aa^\dagger$ do not contribute since
\beq
\label{Maxal}
\langle m| a^\dagger a | n\rangle = \langle m | a a^\dagger | n \rangle =0
\eeq
for $|n\rangle \neq | m\rangle$. As a result, we can drop the second term on the r.h.s.\ of equation \eqref{Maxak}. In fact, this second term originates from the first term on the r.h.s.\ of the expression \eqref{Maxaea}, which is an integrated total derivative in space. This term was not expected to contribute and indeed we see that it does not.

Before proceeding further it is useful to make the following observation.

\vspace{0.5cm}
\noindent
{\bf Parenthesis:} {\it general simplifications in Berry curvature if the energy eigenvalues are not changed.}
~ \vspace{0.3cm}

Assume that the deformations of a Hamiltonian do not alter the eigenvalues (but, potentially alter the eigenvectors). Then, the spectral sum \eqref{berrycurv} for the Berry curvature simplifies. This occurs when the deformations of the Hamiltonian, $H'$, are implemented by similarity transformations $H' = V^{-1} H V$, where $V$ is some invertible, but not necessarly unitary operator. 
Writing $V=\exp[O]$, where $O$ is not necessarily anti-Hermitian, we find that the infinitesimal deformation of the Hamiltonian is
\beq
\label{Maxam}
\delta H = [H,O]~.
\eeq

When we compute the components of the Berry curvature in two directions with the above property \eqref{Maxam} (say, directions 1,2), we find
\beq
\label{Maxan}
(\bF_{12})_{nn'} = \sum_{E_m\neq E_n=E_{n'}} \frac{\langle n| [H,O_1]|m \rangle \langle m | [H,O_2]|n' \rangle}{(E_n-E_m)^2} 
- (1\leftrightarrow2)~.
\eeq
Since $\langle n | [H,O_1]|m \rangle = (E_n-E_m) \langle n|O_1 |m \rangle$ eq.\ \eqref{Maxan} simplifies to
\beq
\label{Maxao}
(\bF_{12})_{nn'} = \sum_{E_m\neq E_n} \langle n| O_1|m \rangle \langle m | O_2 |n'\rangle  - (1\leftrightarrow2)~.
\eeq
Adding and subtracting a contribution from states $|m\rangle$ with $E_m=E_n$, and using the completeness relation $\sum_m |m\rangle \langle m |=1$, we obtain
\beq
\label{Maxap}
(\bF_{12})_{nn'}= \langle n|[O_1,O_2]|n' \rangle~.
\eeq

\hfill
$\blacksquare$

The case of electromagnetism that we consider in this appendix falls directly within the premise of the above parenthesis. Indeed, it is not hard to show that
\beq
\label{Maxaq}
\d_{\g^2} H = [H, O_{\g^2} ]~, ~~
\d_\theta H = [H, O_{\theta}]~,
\eeq
where
\beq
\label{Maxara}
O_{\g^2} = \frac{1}{4\g^2} \sum_{\vec k}\sum_{\epsilon=\pm} 
\left( a_{\vec k,\epsilon} a_{-\vec k,\epsilon} + a^\dagger_{\vec k,\epsilon} a^\dagger_{-\vec k,\epsilon} \right)~,
\eeq
and (after dropping the $a^\dagger a, aa^\dagger$ terms that do not contribute)
\beq
\label{Maxarb}
O_\theta = - \frac{i \g^2}{32 \pi^2} \sum_{\vec k} \sum_{\epsilon=\pm} 
\epsilon \left( a_{\vec k,\epsilon} a_{-\vec k,\epsilon} - a^\dagger_{\vec k,\epsilon} a^\dagger_{-\vec k,\epsilon} \right)~.
\eeq

Then, applying the formula \eqref{Maxap} to the photon external states $|n\rangle$, $|n'\rangle$ we obtain\footnote{To obtain this result we drop a term with alternating signs $\displaystyle \sum_{\vec k} \sum_{\epsilon=\pm} \epsilon \langle n | n'\rangle$. This guarantees that the ground state has vanishing Berry phase, which is a property anticipated to hold in CPT invariant theories.}
\beq
\label{Maxas}
(\bF_{\g^2 \theta})_{nn'} = \langle n | [O_{\g^2},O_\theta ] |n'\rangle~,
\eeq
which can be manipulated further by using the canonical commutation relations of the creation and annihilation operators. After a few steps we arrive at the formula
\beq
\label{Maxat}
(\bF_{\g^2\theta})_{nn'} = - \frac{i}{32\pi^2} \sum_{\vec k} \langle n | (N_{\vec k, +} - N_{\vec k,-}) | n'\rangle~.
\eeq
$N_{\vec k,\epsilon}$ is the number operator at 3-momentum $\vec k$ and helicity $\epsilon$. The r.h.s.\ of \eqref{Maxat} is non-vanishing only when the external states are identical $|n'\rangle = |n\rangle$. We express this fact with a symbolic $\delta$-function $\delta_{n,n'}$. If ${\mathfrak n}_+$, ${\mathfrak n}_-$ are respectively the total number of photons with $+$ or $-$ helicity in the state $|n\rangle$ we finally obtain
\beq
\label{Maxau}
(\bF_{\g^2\theta})_{nn'} = -\frac{i}{32\pi^2} ({\mathfrak n}_+-{\mathfrak n}_-) \delta_{nn'}~.
\eeq
In terms of the complex coupling $\tau = \frac{\theta}{2\pi} + \frac{4\pi i}{\g^2}$ 
\beq
\label{Maxaw}
(\bF_{\tau\bar \tau})_{nn'} = \frac{1}{8} ({\mathfrak n}_+- {\mathfrak n}_-) \frac{1}{({\rm Im}\tau)^2}\delta_{nn'}~.
\eeq

\section{Details and conventions of operator-state correspondence}
\label{conventions}

We follow closely the conventions of Ref.\ \cite{Dolan:2002zh}, where one can find a detailed exposition of the facts listed here. In this brief appendix we focus, for the benefit of the reader, on specific aspects that play a key role in the main text. 

\subsection{Details of conformal algebras}
\label{convN0}

The conformal algrebra on $\IR^{d-1,1}$ involves the generators of translations and special conformal transformations, $P_\mu$, $K_\mu$ $(\mu,\nu=0,1,\ldots, d-1)$ respectively, the Lorentz generators for $SO(d-1,1)$, $M_{\mu\nu}=-M_{\nu\mu}$, and the generator of scale transformations $D$. The commutation relations of these generators are well-known and will not be repeated here. 

In the operator-state correspondence the CFT is Wick rotated to Euclidean signature and quantized radially. Equivalently, with a conformal transformation it is mapped to the hypercylinder $\IR_\tau \times S^d$. Under this map the origin on the plane transforms to $\tau=-\infty$ and the radial infinity on the plane to $\tau=+\infty$.

A local quasi-primary field $\OO_I(x)$ maps to a state $|\OO\rangle_I$ by action of the operators at the origin of $\IR^d$ on the vacuum
\beq
\label{convaa}
| \OO \rangle_I = \OO_I (0) |0\rangle~.
\eeq
The resulting states are conformal primary states satisfying the relations
\beq
\label{convab}
K_a |\OO \rangle_I = 0~, ~~ D|\OO\rangle_I = i \Delta_I |\OO\rangle_I~.
\eeq
$\Delta_I$ is the scaling dimension of the operator $\OO_I$. Similarly, the conjugate fields $\bar\OO_I(x) = \OO_I(x)^\dagger$ define the bra-states
\beq
\label{convac}
_{\bar I} \langle \bar\OO | = \langle 0 | \bar\OO_I(0)~.
\eeq
The states $|\OO\rangle_I$ are not normalizable (which explains why the Hermitian operator $D$ has imaginary eigenvalues on them, \eqref{convab}).

Now comes the main point we want to emphasize. It is convenient to organize the unitary positive energy representations of the theory by going to a new basis where all operators $\OO$ are transformed by the similarity transformation
\beq
\label{convad}
\OO ~\longrightarrow ~ e^{\frac{\pi}{4}(P_0-K_0)} \OO e^{-\frac{\pi}{4}(P_0-K_0)} ~.
\eeq
For example, the transformation of the dilatation operator $D$ is the so-called conformal Hamiltonian $H$, specifically
\beq
\label{convae}
- e^{\frac{\pi}{4}(P_0-K_0)} i D e^{-\frac{\pi}{4}(P_0-K_0)}  = H~.
\eeq
Accordingly, the bra and ket states transform to
\beq
\label{convaf}
| I\rangle = e^{\frac{\pi}{4}(P_0-K_0)} |\OO\rangle_I~, ~~
\langle \bar I | = \, _{\bar I} \langle \bar \OO | e^{\frac{\pi}{4}(P_0-K_0)}~.
\eeq
These states are normalizable positive energy eigenstates of the Hamiltonian
\beq
\label{convag}
H | I\rangle = \Delta_I |I\rangle
\eeq
(see \cite{Dolan:2002zh,Mack:1975je} for further details).

\subsection{Superconformal algrebra of 2d $\NN=(2,2)$ theories}
\label{conv2dN2}

The global left-moving part of the $\NN=(2,2)$ superconformal algebra in the NSNS sector includes the Virasoro generators $L_0,L_{\pm 1}$, the $U(1)_R$ charge $J_0$ and the supercharges $G^-_{\pm 1/2}$, $G^+_{\pm 1/2}$. There is a similar right-moving copy of these generators. On $\IR^2$ (with complex coordinates $(z,\bar z)$) the momentum component $P_0=-\d_z-\d_{\bar z}=L_{-1}+\bar L_{-1}$ and the component of the special transformation generator $K_0$ is likewise $K_0=-z^2 \d_z-\bar z^2 \d_{\bar z} = L_{1}+\bar L_{1}$. The dilatation operator is $iD = -i z \d_z -i \bar z \d_{\bar z}= i \left( L_0 +\bar L_0\right)$.

Applying the similarity transformation \eqref{convad} to the above supercharges we obtain the calligraphic generators $\LL_0, \LL_{\pm 1}, \JJ_0, \QQ^\pm , \SS^\pm$ with 
\begin{subequations}
\bea
\label{conv2dN2aa}
\QQ^+ &=& e^{\frac{\pi}{4}(L_{-1}-L_1)} G^+_{-1/2} e^{-\frac{\pi}{4}(L_{-1}-L_1)} 
\\
\QQ^- &=& e^{\frac{\pi}{4}(L_{-1}-L_1)} G^-_{+1/2} e^{-\frac{\pi}{4}(L_{-1}-L_1)} 
\\
\SS^- &=& e^{\frac{\pi}{4}(L_{-1}-L_1)} G^+_{+1/2} e^{-\frac{\pi}{4}(L_{-1}-L_1)} 
\\
\SS^+ &=& e^{\frac{\pi}{4}(L_{-1}-L_1)} G^-_{-1/2} e^{-\frac{\pi}{4}(L_{-1}-L_1)} 
~.
\eea
\end{subequations}
Obviously, the similarity transformation can be performed separately on the left- and right-movers. Hermitian conjugation operates as follows
\beq
\label{conv2dN2ab}
\left( \QQ^+ \right)^\dagger = \QQ^- ~, ~~ \left( \SS^+ \right)^\dagger = \SS^-
~.
\eeq
Some of the (anti)-commutation relations of interest for the left-moving generators (similar relations apply to the right-movers) are
\begin{subequations}
\bea
\label{conv2dN2ac}
&&\left[ \LL_m, \LL_n \right] = (m-n) \LL_{m+n}~, ~~ \left[ \LL_m , \JJ_0 \right] =0~, ~~
m, n=\pm 1, 0~,
\\
\label{conv2dN2acb}
&&\{ \QQ^-, \QQ^+ \} = 2\LL_0-\JJ_0 ~, ~~ \{ \SS^-, \SS^+ \} = 2\LL_0+\JJ_0~,
\\
&&\{ \QQ^-, \SS^- \} = 2 \LL_1 ~, ~~ \{ \QQ^+, \SS^+ \} = 2 \LL_{-1}~, ~~ 
\{ \QQ^-, \SS^+\} = 0~, ~~ \{ \QQ^+, \SS^-\} = 0~,
\\
&& [ \LL_0, \QQ^\pm ] = \pm \frac{1}{2} \QQ^\pm~, ~~ [ \LL_0, \SS^\pm ]= \pm \frac{1}{2} \SS^\pm,~~
\\
&& [ \JJ_0 , \QQ^\pm ] = \pm \QQ^\pm ~, ~~ [\JJ_0, \SS^\pm ]= \mp \SS^\pm~.
\eea
\end{subequations}
The Hamiltonian operator and the total $U(1)_R$ charge are the sum of their left- and right-moving counterparts
\beq
\label{conv2dN2ad}
H = \LL_0+\bar \LL_0 ~, ~~
R = \JJ_0 + \bar \JJ_0~.
\eeq

\subsection{Superconformal algebra of 4d $\NN=2$ theories}
\label{conv4dN2}

The superconformal algebra of 4d $\NN=2$ theories includes the real supercharges $Q^i_\alpha$, $\bar Q_{i\dot\alpha}$ and their superconformal partners $S_i^\alpha$, $\bar S^{i\dot\alpha}$. The indices $i=1,2$ are $SU(2)_R$ indices and the indices $(\alpha,\dot\alpha= \pm)$ are standard spinor indices. These supercharges realize the $SU(2,2|2)$ Lie superalgebra. 

Applying the similarity transformation \eqref{convad} to the above supercharges we obtain the calligraphic generators \cite{Minwalla:1997ka,Dolan:2002zh}
\begin{subequations}
\label{SCAconventions}
\bea
\label{conv4dN2aa}
\QQ^{+i}_\alpha &=& e^{\frac{\pi}{4}(P_0-K_0)} Q^i_\alpha e^{-\frac{\pi}{4}(P_0-K_0)} 
= \frac{1}{\sqrt 2} \left( Q^i_\alpha + \sigma_{0\, \alpha\dot\alpha} \bar S^{i\dot\alpha} \right)~,
\\
\QQ^{-\alpha}_i &=& e^{\frac{\pi}{4}(P_0-K_0)} S_i^\alpha e^{-\frac{\pi}{4}(P_0-K_0)}  
= \frac{1}{\sqrt 2} \left( S_i^\alpha + \bar Q_{i\dot\alpha} \bar \sigma_0^{~\dot\alpha \alpha} \right)~,
\\
- \SS^{-i\dot\alpha} &=& e^{\frac{\pi}{4}(P_0-K_0)} \bar S^{i\dot\alpha} e^{-\frac{\pi}{4}(P_0-K_0)} 
=\frac{1}{\sqrt 2} \left( \bar S^{i\dot\alpha} - \bar\sigma_0^{~\dot\alpha \alpha} Q^i_\alpha \right)~,
\\
\label{conv4dN2aaa}
\SS^+_{i\dot\alpha} &=& e^{\frac{\pi}{4}(P_0-K_0)} \bar Q_{i\dot\alpha} e^{-\frac{\pi}{4}(P_0-K_0)} 
=\frac{1}{\sqrt 2} \left( \bar Q_{i\dot\alpha} - S_i^\alpha \sigma_{0 \, \alpha \dot\alpha} \right)~,
\eea
\end{subequations}
which play a key role in the main text. These operators obey (among other things) the (anti)commutation relations
\begin{subequations}
\bea
\label{conv4dN2ab}
\{ \QQ^{+i}_{\alpha}, \QQ^{-\beta}_j \} &=& 2 \delta^i_{~j}\delta_{\alpha}^{~\beta} H 
+ 4 \delta^i_{~j} \tilde M_\alpha^{~\beta}
- 4 \delta_{\alpha}^{~\beta} R^i_{~j}~,
\\
\{ \SS^{-i\dot\alpha}, \SS^+_{j\dot\beta} \} &=&  2 \delta^i_{~j}\delta^{\dot \alpha}_{~\dot \beta} H 
- 4 \delta^i_{~j} \tilde {\bar M}^{\dot \alpha}_{~\dot \beta}
+ 4 \delta^{\dot \alpha}_{~\dot \beta} R^i_{~j}~,
\eea
\end{subequations}
\beq
\label{conv4dN2ac}
[H, \QQ^\pm ]=\pm \frac{1}{2} \QQ^\pm ~, ~~
[H, \SS^\pm ] =\pm \frac{1}{2} \SS^\pm
\eeq
with Hermiticity properties
\beq
\label{conv4dN2ad}
\left( \QQ^{+i}_\alpha \right)^\dagger = \QQ^{-\beta}_i \sigma_{0\beta \dot \alpha}~, ~~
\left( \SS^+_{i\dot\alpha} \right)^\dagger = \sigma_{0\alpha\dot\beta} \SS^{-i\dot\beta}~.
\eeq
We assume by convention that $\sigma_0 = \bar \sigma_0= 1$. The rotation generators $\tilde M_{\alpha}^{~\beta}$, $\tilde{\bar M}^{\dot \alpha}_{~\dot\beta}$ are defined as the transformation \eqref{convad} of the generators
\beq
\label{conv4dN2ae}
M_\alpha^{~\beta} = -\frac{i}{4}\left( \sigma^\mu \bar\sigma^\nu \right)_\alpha^{~\beta} M_{\mu\nu}~, ~~
\bar M^{\dot \alpha}_{~\dot \beta} = -\frac{i}{4}\left( \bar \sigma^\mu \sigma^\nu \right)^{\dot \alpha}_{~\dot \beta} M_{\mu\nu}~.
\eeq

The $\NN=2$ $U(2)_R$-symmetry generators are
\beq
\label{conv4dN2af}
\left( R^i_{~j} \right) = \left( {R_3 \atop R_-} ~ {R_+ \atop -R_3} \right) -\frac{R}{4} \left( {1 \atop 0 } ~ {0 \atop 1} \right)~,
\eeq
where $R_\pm, R_3$ are $SU(2)_R$ generators and $R$ is a $U(1)_R$ generator normalized so that chiral primary operators obey the scaling dimension relation $\Delta=\frac{R}{2}$. The $U(1)_R$ charges of the supercharges are
\beq
\label{conv4dN2ag}
\left[ R, \QQ^\pm \right] = \mp \QQ^\pm ~, ~~
\left[ R, \SS^\pm \right] = \pm \SS^\pm~.
\eeq

\section{Technical results in $2d$ $\NN=(2,2)$ SCFTs}
\label{2dtechnical}

In this appendix we evaluate the contact term $\RR$ defined in eq.\ \eqref{2dba}. We perform the computation on the cylinder where the integrated insertions are evaluated by default at equal Euclidean time $\tau =0$. Since any potential contribution is expected to arise when the operators $\varphi_k$ and $\bar\varphi_{\bar l}$ come close together, we can evaluate it by invoking the OPE of these operators. Moreover, since the amplitude $\RR$ depends only on the relative distance of the $\varphi$ insertions we can also recast it in the form
\beq
\label{2dbb}
\RR = \frac{1}{2\pi} \int_{-\pi}^\pi d\theta \, \langle \bar J | \Big[ 
\varphi_k(0,\theta) \bar\varphi_{\bar l}(0,0) - \bar\varphi_{\bar l}(0,\theta) \varphi_k(0,0) \Big] |I\rangle~.
\eeq

Now consider the general OPE of a scaling dimension $\left( \frac{1}{2},\frac{1}{2} \right)$ chiral primary field $\varphi$ with a $\left( \frac{1}{2},\frac{1}{2} \right)$ anti-chiral primary $\bar\varphi$ on the cylinder. On the Euclidean plane with complex coordinates $(z,\bar z)$ we have 
\beq
\label{2dbc}
\varphi(z_1,\bar z_1) \bar\varphi(z_2,\bar z_2) =
\sum_\rho \frac{D_{\phi\bar\phi}^\rho \OO_\rho(z_2,\bar z_2)}
{z_{12}^{1-h_\rho} \bar z_{12}^{1-\bar h_\rho}}~.
\eeq
The operator $\OO_\rho$ has left-right scaling dimensions $(h_\rho,\bar h_\rho)$. When we transform from the plane to the cylinder with the change of coordinates
\beq
\label{2dbd}
z= e^{-i w} = e^{-i (\theta+i \tau)}
\eeq
the $(\frac{1}{2},\frac{1}{2})$ (anti)chiral primaries transform as
\beq
\label{2dbe}
\varphi(z,\bar z) = e^{\frac{i}{2} (w-\bar w)} \varphi(w,\bar w)~.
\eeq
In addition,
\beq
\label{2dbf}
\OO_\rho(z,\bar z) = i^{h_\rho - \bar h_\rho} 
e^{i h_\rho w - i \bar h_\rho \bar w} \OO_\rho (w,\bar w)~,
\eeq
\beq
\label{2dbg}
z_{12}^{h_\rho-1} \bar z_{12}^{\bar h_\rho -1}
= e^{-i (h_\rho-1) w_2 + i (\bar h_\rho-1)\bar w_2} \left( -1 + e^{-i w_{12}} \right)^{h_\rho-1}
\left( -1 + e^{i \bar w_{12}} \right)^{\bar h_\rho-1}~.
\eeq
Inserting these formulae into \eqref{2dbc} we obtain on the Euclidean cylinder the OPE
\bea
\label{2dbi}
&&\varphi(w_1,\bar w_1) \bar\varphi(w_2,\bar w_2) =
\cr
&&=\sum_\rho i^{h_\rho -\bar h_\rho} D_{\phi\bar\phi}^\rho 
e^{-\frac{i}{2} w_{12} + \frac{i}{2} \bar w_{12}}
\left( -1 + e^{-i w_{12}} \right)^{h_\rho-1}
\left( -1 + e^{i \bar w_{12}} \right)^{\bar h_\rho-1}
\OO_\rho (w_2,\bar w_2)~.
\eea
As $w_{12}\to 0$ the only terms that are singular in this OPE arise from the identity and $(1,0)$, $(0,1)$ operators. Expanding the exponentials and keeping at most the linear terms in the expansion (which is sufficient for our purposes) we obtain
\bea
\label{2dbj}
&&\varphi(w_1,\bar w_1) \bar\varphi(w_2,\bar w_2) 
=\sum_\rho  D_{\phi\bar\phi}^\rho 
\frac{1-\frac{i}{2} h_\rho w_{12} +\frac{i}{2} \bar h_\rho \bar w_{12}}
{w_{12}^{1-h_\rho} \bar w_{12}^{1-\bar h_\rho}}
\OO_\rho (w_2,\bar w_2)
\cr
&&\simeq D^{1}_{\phi\bar\phi} 
\frac{1}{|w_{12}|^2} 
+D^{(1,0)}_{\phi\bar\phi} \left( \frac{1}{\bar w_{12}} - \frac{i}{2} \frac{w_{12}}{\bar w_{12}} \right) 
\OO_{(1,0)}(w_2,\bar w_2)
\cr
&&+D^{(0,1)}_{\phi\bar\phi} \left( \frac{1}{w_{12}} + \frac{i}{2} \frac{\bar w_{12}}{w_{12}} \right) 
\OO_{(0,1)}(w_2,\bar w_2)
+\ldots~.
\eea

Following the discussion below equation \eqref{4darh}
we insert \eqref{2dbj} into the modified version of \eqref{2dbb} 
\beq
\label{2dbka}
\RR = \frac{1}{2\pi} \int_{-\pi}^\pi d\theta \, \langle \bar J | \Big[ e^{\tau_1-\tau_2}
\varphi_k(\tau_1,\theta) \bar\varphi_{\bar l}(\tau_2,0) - e^{\tau_2-\tau_1} \bar\varphi_{\bar l}(\tau_1,\theta) \varphi_k(\tau_2,0) \Big] |I\rangle
\eeq
displacing in Euclidean time (or imaginary Minkowski time). We set
\beq
\label{2dbk}
w_1= \theta -\frac{i}{2} \varepsilon~, ~~ w_2 =  \frac{i}{2}\varepsilon ~~
\Rightarrow ~~ w_{12} = \theta - i\varepsilon
\eeq
for the first term on the r.h.s.\ of \eqref{2dbka} and 
\beq
\label{2dbl}
w_1= \frac{i}{2} \varepsilon~, ~~ w_2 = \theta - \frac{i}{2}\varepsilon ~~
\Rightarrow ~~ w_{12} = -\theta + i\varepsilon
\eeq
for the second. We obtain
\bea
\label{2dbm}
\RR &=& - \frac{1}{\pi} g_{I\bar J} \int_{-\pi}^\pi d\theta \, \left[
D^1_{k\bar l} \frac{\varepsilon}{\theta^2 +\varepsilon^2} 
+q D_{k\bar l}^{(1,0)} \frac{-i}{\theta+i\varepsilon}
+\bar q D_{k\bar l}^{(0,1)} \frac{i}{\theta-i\varepsilon} \right]
\cr
&=& - g_{I\bar J} \left[ D^1_{k\bar l} - q D_{k\bar l}^{(1,0)} -  \bar q D_{k\bar l}^{(0,1)} \right]~.
\eea
Using 
\beq
\label{2dbn}
D^1_{k\bar l} = g_{k\bar l} ~, 
\eeq
\beq
\label{2dbo}
D^{(1,0)}_{k\bar l} =\frac{3}{c} g_{k\bar l} ~, 
\eeq
\beq
\label{2dbp}
D^{(0,1)}_{k\bar l} =\frac{3}{c} g_{k\bar l} 
\eeq
we finally deduce that
\beq
\label{2dbq}
\RR = - \left( 1 - \frac{3}{c} (q+\bar q) \right)
g_{I \bar J} g_{k\bar l}~.
\eeq
We also used the integrals
\beq
\label{2dbr}
\lim_{\varepsilon \to 0} \int_{-\pi}^\pi  d\theta~ \frac{\varepsilon}{\theta^2 +\varepsilon^2} = \pi~,
\eeq
\beq
\label{2dbs}
\lim_{\varepsilon \to 0} \int_{-\pi}^\pi d\theta ~ \frac{1}{\theta \pm i\varepsilon} =\mp \pi i 
\eeq
and in the second line of \eqref{2dbm} the identities
\beq
\label{2dbt}
\langle J | \OO_{(1,0)} ( w=0) | I \rangle = i q g_{I\bar J}~,~~
\langle J | \OO_{(0,1)} ( w=0) | I \rangle =- i \bar q g_{I\bar J}~.
\eeq

\end{appendix}

\newpage

\bibliography{ttstarpaper}
\bibliographystyle{utphys}

\end{document}